\documentclass[preprint,12pt]{elsarticle}

\usepackage{amssymb}
\usepackage{amsmath}

\usepackage{subfigure}
\usepackage{subfig}
\usepackage[acronym]{glossaries}

\newacronym{cfd}{CFD}{computational fluid dynamics}
\newacronym{des}{DES}{detached eddy simulation}
\newacronym{pdf}{PDF}{probability density function}
\newacronym{les}{LES}{large eddy simulation}
\newacronym{rans}{RANS}{Reynolds-averaged Navier-Stokes equations}
\newacronym{lre}{LREs}{liquid rocket engines}
\newacronym{sbli}{SBLI}{shock wave/boundary layer interaction}
\newacronym{rss}{RSS}{restricted shock separation}
\newacronym{fss}{FSS}{free shock separation}
\newacronym{ddes}{DDES}{delayed detached eddy simulation}

\journal{Aerospace Science and Technology}

\begin{document}

\begin{frontmatter}



\title{Experimental and Numerical Analysis of the Intermittentency in a Nozzle Overexpanded-Flow}


\author{Emanuele Martelli\footnote[1]{emanuele.martelli@polito.it}} 
\affiliation{organization={Dipartimento di Ingegneria Meccanica e Aerospaziale, Politecnico di Torino},
            addressline={Corso Duca degli Abruzzi 24}, 
            city={Torino},
            postcode={10128},
            country={Italy}} 

\author{Vincent Jaunet} 
\affiliation{organization={ISAE-ENSMA Institut PPRIME},
            addressline={Avenue Clement Ader}, 
            city={Chasseneuil-du-Poitou},
            postcode={86360},
            country={France}} 

\author{Giacomo Della Posta} 
\affiliation{organization={ESA - ESRIN},
            addressline={via Galileo Galilei 1}, 
            city={Frascati},
            postcode={00044},
            country={Italy}} 

\author{Matteo Bernardini} 
\affiliation{organization={Sapienza University of Rome},
            addressline={Via Eudossiana 18}, 
            city={Roma},
            postcode={00184},
            country={Italy}} 

\begin{abstract}
  The present work reports an investigation into the statistical properties of wall-pressure fluctuations in a highly over-expanded nozzle flow, characterized by significant shock-induced flow separation. This regime is extremely hazardous to rocket nozzles, as it leads to very high off-axis loads. The database under investigation has been obtained both experimentally and numerically by means of a hybrid RANS/LES simulation of the flow issuing from a sub-scale Truncated Ideal Contour (TIC) nozzle, fed with cold air and operating at a Reynolds number on the order of $10^6$. The experimental campaign was conducted in the S150 supersonic wind tunnel at the Institut PPRIME in Poitiers.
  The degree of over-expansion is quantified by the nozzle pressure ratio (NPR).  Pressure fluctuations are extracted from several probes positioned along the nozzle wall, considering different NPR values. The intermittent behavior is investigated using conditional statistics based on the wavelet transform, which demonstrates that the aerodynamic loads of the over-expanded jet consist of intermittent bursts rather than continuous variations. The wavelet analysis reveals scale-by-scale intermittency and, in particular, shows that the wall-pressure signals exhibit a significant degree of intermittency around the frequency associated with aerodynamic side-loads. The statistics of these intermittent events, in terms of the time delay between occurrences and in terms of their amplitude, are found to be weakly sensitive to NPRs and to locations along the nozzle wall and appear to follow a universal behaviour that can be modelled by a log-normal distribution. This finding may support the development of a stochastic model of the aerodynamic side-loads.
\end{abstract}




\end{frontmatter}



\section{Introduction}

Supersonic nozzles that equip \gls{lre} of launchers' first stages experience internal flow separation during the ignition sequence, since they are necessarily over-expanded during this transient.
Internal flow separation entails \gls{sbli} and consequent thermal and fatigue loading, mainly due to off-axis side loads that are intense enough to induce the structural failure of the nozzle~\cite{Nave1973SeaLS}. Depending on the nozzle contour and on the nozzle pressure ratio, defined as the ratio between the inlet total pressure and the ambient pressure, $NPR = p_0/p_a$, two different kinds of shock-induced flow separation can appear: \gls{fss} and \gls{rss}~\cite{Nave1973SeaLS}.
Earlier investigations considered the problem by developing analytical and empirical tools capable of describing the occurrence of side loads, originating from the \gls{fss} state, the \gls{rss} state, or the transition from one state to another~\cite{schmucker1984side,dumnov1996unsteady}. All of these methods were mainly intended for use as design tools.  

As both numerical and experimental disciplines have made numerous advances in the last decades, efforts have recently shifted towards a more physical understanding of the mechanisms responsible for off-axis forces in over-expanded nozzles with internal flow-separation.
The experimental work of Baars et al.~\cite{baars2012} on a parabolic nozzle was one of the first to conduct a Fourier-azimuthal decomposition of the unsteady wall-pressure field, revealing how the breathing zero-th mode is characterized by most of the total resolved pressure-fluctuation energy, whereas the first azimuthal mode, inducing side loads, has less than half of the total energy. 
Later, the experimental work of Jaunet et al.~\cite{jaunet2017} on a truncated ideal contour (TIC) nozzle,  confirmed that the low-frequency shock motion is mainly linked to the axisymmetric azimuthal mode, whereas the middle-frequency peaks identified in the spectral analysis are shown to be very organized in the azimuthal direction, each one of them clearly associated with a preferred azimuthal mode. At NPR=9, the strongest peak is entirely contained in the first azimuthal mode, which is the only mode that can contribute to the off-axis loads. 
The authors have also shown that this tonal dynamics was not due to transonic resonance~\cite{Loh_2002}, but it was not yet clear whether it could be attributed to a screech-like mechanism as it did not radiate tonal sound, although its frequency scaled nicely with the screech frequency model prediction~\cite{jaunet2017}.
Martelli et al.~\cite{Martelli_jfm_20} conducted one of the first \gls{ddes} on a similar nozzle, and confirmed the presence of a coherent azimuthal anti-symmetric mode. In their work, the authors also proposed a model for a screech-like resonance, consisting of a closed feedback loop. The forward path involves the shedding of vorticity from the triple point of the Mach disk toward the second Mach disk in the plume, while the backward path involves the subsonic recirculation region between the nozzle-wall and the detached supersonic shear layer.
For the backward path, Bakulu et al.~\cite{Bakulu_jfm_21} instead proposed one of the guided jet modes~\cite{Tam_Hu_1989}, whereas Morisco et al.~\cite{Morisco_23} again indicated the subsonic recirculation region. Jaunet and Lehsnasch~\cite{Jaunet_Lehnasch_2025} have recently shown that a supersonic over-expanded jet, such as the one exiting a TIC nozzle, is an excellent support for a variety of guided jet modes that could close the feedback loop. Nevertheless, since no frequency prediction and detailed observation of these modes are available, the exact nature of the feedback loop in this case remains an open question and deserves further investigation.
One aspect on which there seems to be agreement, instead, is that the entire jet column is subject to  unsteadiness, observable in various azimuthal modes frequencies, and a particular characteristic of this unsteady behaviour is its intermittent nature. 
In the work of Camussi et al.~\cite{CAMUSSI20171} and Kearney et al.~\cite{kearney_2013}  it is found that both the external field and the nozzle-wall pressure signature emanating from a  compressible jet are characterized indeed by a high degree of intermittency. This term, in statistics, denotes the occurrence of spiky events which are sporadic and burst-like.  It may happen indeed, that non-linear interactions occasionally concur in such a way that responses of very large amplitudes are observed.
Such statistical deviations are commonly referred to as extreme events.
These instances cause higher-order moments to converge with greater difficulty, suggesting a significant departure from Gaussian statistics and hence non-homogeneous distribution of energy in time.
In this paper, we aim to characterise, for the first time to the authors' knowledge, the intermittency of the wall-pressure signature in the separated region of an over-expanded TIC nozzle, experimentally tested at the PPRIME institute~\cite{jaunet2017} at different NPRs and numerically reproduced by means of a \gls{ddes} at NPR = 9, which corresponds to the NPR of maximum aerodynamic side loads~\cite{jaunet2017}. The final goal is to find the basis for the formulation of a stochastic model capable of describing the intermittent behaviour of the aerodynamic loads.
 
\section{Experimental and numerical test case}

\begin{figure}[t]
     \centering
     \includegraphics[width=0.6\textwidth]{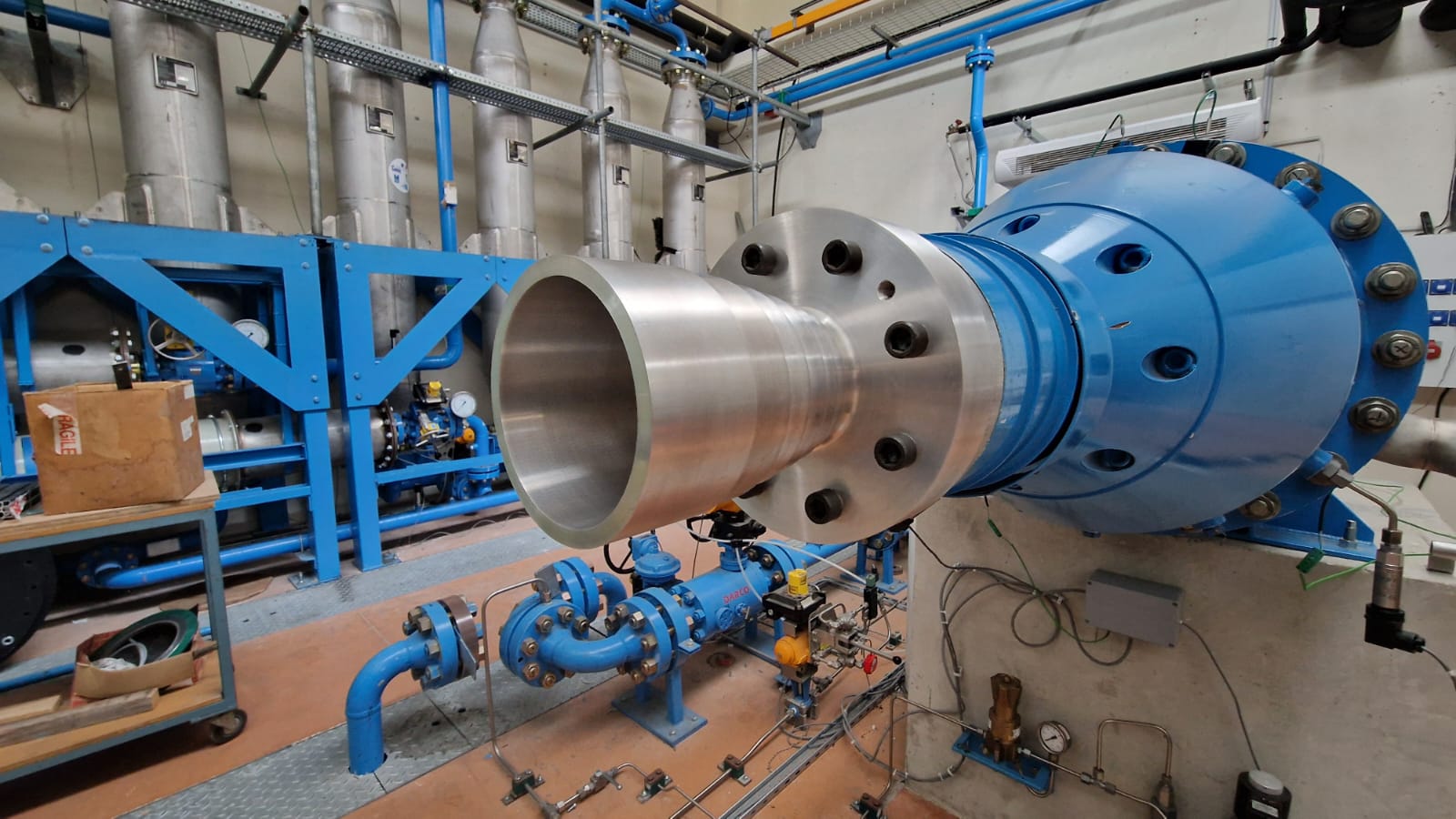} 
     \caption{Experimental apparatus at PPRIME Institute.}
     \label{fig:exp_tic}
\end{figure}

The experimental campaign has been conducted in the S150 supersonic wind tunnel at PPRIME Institute. A rigid TIC nozzle, shown in figure~\ref{fig:exp_tic}, is considered with a divergent length $L = 0.1827$ m, a throat diameter $D_t$ = 0.038 m and an exit diameter $D_e$ = 0.097 m, which lead to a full-flowing flow condition with a Mach number
$M$ = 3.5. The nozzle is supplied with cold (total temperature around 260 K) and dry high-pressure air. The inflow stagnation pressure is varied in order to obtain nozzle pressure ratios in the range 6-12. The nozzle Reynolds number range, evaluated assuming the throat radius as the reference length, the total pressure $p_0$, the total temperature $T_0$ and the molecular viscosity taken at the stagnation-chamber condition $\mu_0 = \mu (T_0)$ is:

\begin{equation}
    2.67\cdot 10^6 \le Re =\frac{\sqrt{\gamma}}{\mu_0}\frac{p_0 r_t}{\sqrt{R_{air} T_0}}\le 5.34 \cdot 10^6.
\end{equation}

The nozzle is equipped with 18 flush-mounted Kulite XCQ-062 pressure transducers, distributed along 3 rings of 6 transducers placed equidistantly along the circumference. Additional details on the experimental setup can be found in Jaunet et al.\cite{jaunet2017}.

The over-expanded jet has also been numerically reproduced by carrying out a \gls{ddes} with an in-house code developed at Sapienza University of Rome, previously employed for similar flow cases~\cite{Martelli_jfm_20, martelli_aiaa_17, martelli2019_aiaa,  Cimini21}.
Simulating the unsteady flow features of supersonic nozzle flows such as the one under consideration is particularly challenging: sufficiently high resolution is required both near the walls---to capture attached boundary layers---and further downstream---where unsteady turbulent structures develop. This results in very restrictive time steps, while long simulation times are necessary to capture the expected low-frequency and mid-frequency features of the integrated wall-pressure forces.
For this reason, the \gls{ddes} method proposed by Spalart~\cite{spalart06} has been adopted, as it enables effective resolution of the relevant scales of the separated flow while keeping the computational cost within practical limits.

We solve the three-dimensional Navier-Stokes equations for a compressible, viscous, heat-conducting gas

\begin{equation}
\begin{aligned}
   \frac{\partial \rho}{\partial t} + \frac{\partial (\rho \, u_j)}{\partial x_j}
     & = 0, \\
   \frac{\partial (\rho \, u_i)}{\partial t} + \frac{\partial (\rho \, u_i u_j)}{\partial x_j} +
       \frac{\partial p}{\partial x_i} - \frac{\partial \tau_{ij}}{\partial x_j} & = 0, \\
   \frac{\partial (\rho \, E)}{\partial t} + \frac{\partial (\rho \, E u_j + p u_j)}{\partial x_j}
     - \frac{\partial (\tau_{ij} u_i - q_j)}{\partial x_j} & = 0, 
     \label{eq:ns}
 \end{aligned}
\end{equation}

where $\rho$ is the density, $u_i$ is the velocity component in the $i$-th coordinate direction ($i=1,2,3$),
$E$ is the total energy per unit mass, $p$ is the thermodynamic static pressure.
The total stress tensor $\tau_{ij}$ is the sum of the viscous and the Reynolds stress tensor, 
\begin{equation}
 \label{eq:stress}
 \tau_{ij} = 2 \, \rho \left ( \nu + \nu_t \right ) S^*_{ij} \qquad S^*_{ij} = S_{ij} - \frac 13 \, S_{kk} \, \delta_{ij},
\end{equation}
where the Boussinesq hypothesis is applied through the introduction of the eddy viscosity $\nu_t$, $S_{ij}$ is the strain-rate tensor and $\nu$ the kinematic viscosity, depending on temperature $T$ through Sutherland's law.
Similarly, the total heat flux $q_j$ is the sum of a molecular and a turbulent contribution

\begin{equation}
 q_j = -\rho \, c_p \left ( \frac{\nu}{\mathrm{Pr}} + \frac{\nu_t}{\mathrm{Pr}_t} \right ) \frac{\partial T}{\partial x_j},
\end{equation}

where $\mathrm{Pr}$ and $\mathrm{Pr}_t$ are the molecular and turbulent Prandtl numbers, assumed to be
0.72 and 0.9, respectively.

The \gls{ddes} approach adopted is the one based on the Spalart-Allmaras (SA) turbulence model~\cite{spalart06}, which solves a transport equation for the pseudo eddy viscosity $\tilde{\nu}$:

\begin{multline}
 \label{eq:sa}
 \frac{\partial (\rho \tilde{\nu})}{\partial t} + \frac{\partial (\rho \, \tilde{\nu} \,u_j)}{\partial x_j} = 
 c_{b1} \tilde{S} \rho \tilde{\nu} +
 \frac{1}{\sigma}
 \left [
  \frac{\partial}{\partial x_j} \left [ \left ( \rho \nu + \rho \tilde {\nu} \right ) \frac{\partial \tilde{\nu}}{\partial x_j} \right ] + \right. \\ 
  \left. +  c_{b2} \, \rho \left ( \frac{\partial \tilde{\nu}}{\partial x_j} \right )^2
 \right ]
  -c_{w1}  f_w \rho \left ( \frac{\tilde{\nu}}{\tilde{d}} \right )^2,
\end{multline}

where $\tilde{d}$ is the model length scale, $f_w$ is a near-wall damping function, $\tilde{S}$
a modified vorticity magnitude, and $\sigma,~c_{b1},~c_{b2},~c_{w1}$ model constants.
The eddy viscosity in Eq.~\ref{eq:stress} is related
to $\tilde{\nu}$ through $\nu_t = \tilde{\nu} \, f_{v1}$, where $f_{v1}$ is a
correction function designed to guarantee the correct boundary-layer behaviour in the near-wall region.
The destruction term in Eq.~\ref{eq:sa} is built in such a way that the model reduces to pure \gls{rans} model in attached boundary layers and to a \gls{les} sub-grid scale model in separated flow regions.
This task is accomplished by defining the length-scale $\tilde{d}$ as
\begin{equation}
 \tilde{d} = d_w - f_d \, \textrm{max} \left (0, d_w-C_{DES} \, \Delta \right),
\end{equation}
where $d_w$ is the distance from the nearest wall, $\Delta$ is the subgrid length-scale that controls the wavelengths resolved in LES mode and $C_{DES}$ is a calibration constant equal to 0.20. 
The function $f_d$, designed to be $0$ in boundary layers and $1$ in LES regions, is defined as
\begin{equation}
 \label{eq:sh}
 f_d = 1-\tanh{\left [ \left ( 16 r_d \right )^3 \right ]}, \qquad r_d = \frac{\tilde{\nu}}{k^2 \, d_w^2 \, \sqrt{U_{i,j} U_{i,j}}},
\end{equation}
where $U_{i,j}$ is the velocity gradient and $k$ the von Karman constant.
The introduction of $f_d$ guarantees that the boundary layer is modeled in RANS mode also in the presence of particularly fine grids, for which the spacing in the wall-parallel directions does not exceed the boundary layer thickness. This precaution is needed to prevent the phenomenon of model stress depletion\cite{deck12}.
The sub-grid length scale in this work is defined according to Deck ~\cite{deck12}, and it depends on the flow itself, through $f_d$ as

\begin{equation}
 \label{eq:delta}
 \Delta = \frac{1}{2} 
 \left [
 \left (1+\frac{f_d-f_{d0}}{|f_d-f_{d0}|} \right ) \, \Delta_{\textrm{max}} +
 \left (1-\frac{f_d-f_{d0}}{|f_d-f_{d0}|} \right ) \, \Delta_{\textrm{vol}}
 \right ],
\end{equation}

where $f_{d0} = 0.8$, $\Delta_{\mathrm{max}} = \max (\Delta x, \Delta y, \Delta z)$ and $\Delta_{\mathrm{vol}} = (\Delta x \cdot \Delta y \cdot \Delta z)^{1/3}$.
The main idea of this formulation is to take advantage of the $f_d$ function to switch
between $\Delta_{max}$, needed to shield the boundary layer, and $\Delta_{\mathrm{vol}}$, needed to ensure a rapid destruction of modelled viscosity and trigger the formation of turbulent structures.

The computational domain includes the nozzle and the external ambient: starting from the nozzle throat, the outflow boundary extends for 150 throat radii in the longitudinal direction and for 76 throat radii in the radial direction, relative
to the nozzle axis. The topology of this computational domain is the same as the one used by Martelli et al.~\cite{Martelli_jfm_20}.  
The total pressure, total temperature, and flow direction are imposed at the nozzle inflow, while an assigned ambient pressure $p_a$ is prescribed at the outer boundaries.  
The nozzle walls are treated with a no-slip adiabatic boundary condition.  
The final grid consists of approximately 115 million cells, including 260 cells in the azimuthal direction. The mesh resolution and total number of cells are based on previous similar simulations~\cite{Martelli_jfm_20,Cimini21}.

\section{Flow field organization and spectral characterization}

\begin{figure}[!ht]
     \centering
          \subfigure[]{\includegraphics[width=0.60\textwidth]{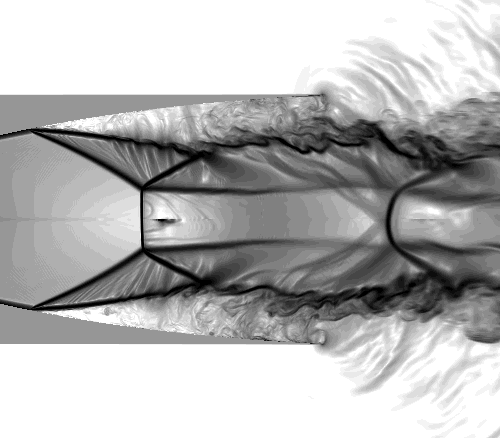}} \\
         \subfigure[]{\includegraphics[width=0.45\textwidth]{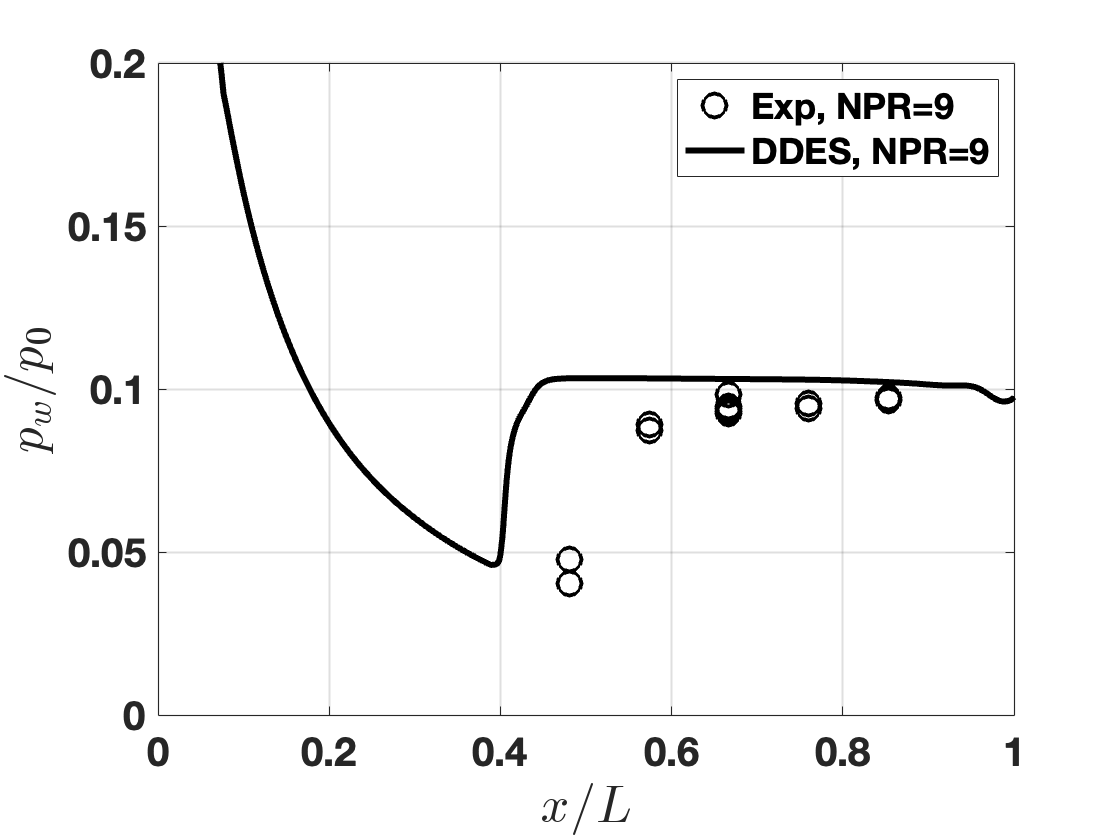}}
          \subfigure[]{\includegraphics[width=0.45\textwidth]{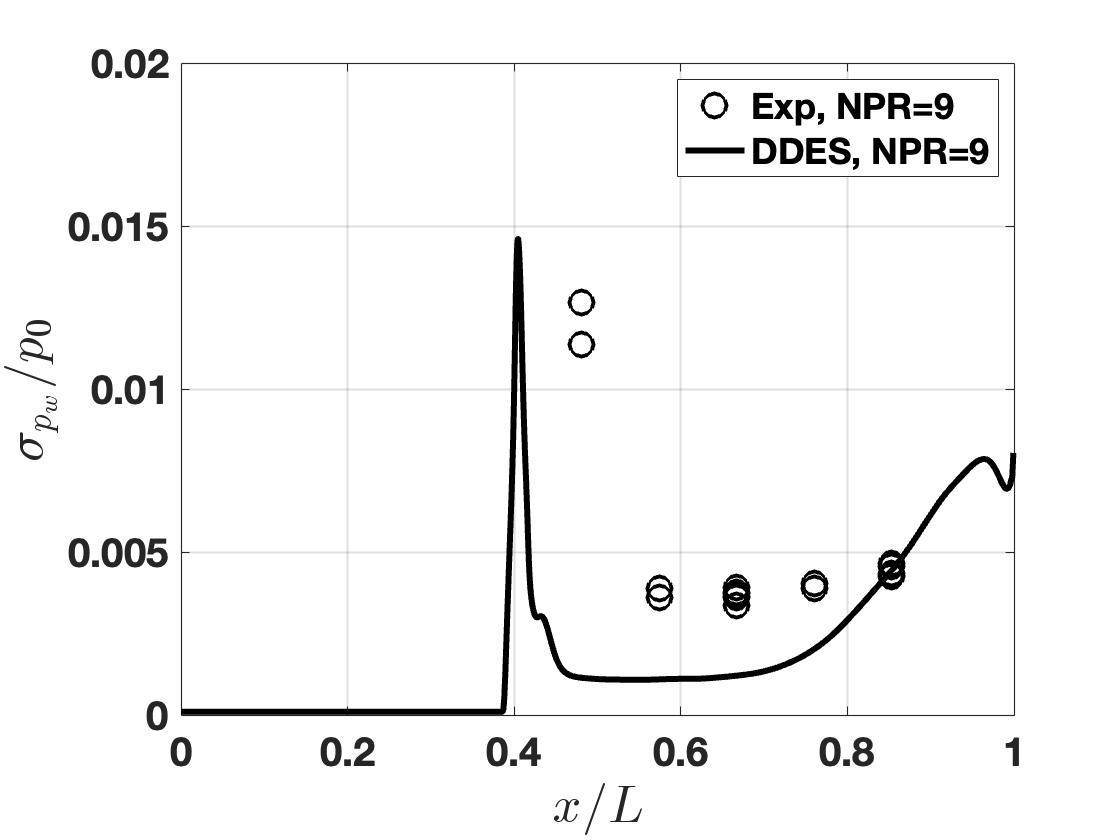}} 
     \caption{a) Instantaneous visualisation of the turbulent and shock structures at NPR=9 from \gls{ddes}: numerical Schlieren on a longitudinal view; 
     b) experimental and numerical mean wall-pressure; 
     c) experimental and numerical standard deviation of the wall-pressure signals along the streamwise direction.}
     \label{fig:ddes_tic}
\end{figure}

Figure~\ref{fig:ddes_tic}a shows an instantaneous numerical Schlieren image in a longitudinal view, illustrating an enlarged visualization of the flow topology under consideration. First, the separation-induced converging conical shock is observed, which reflects as a Mach disk on the nozzle centreline, followed by a diverging conical shock originating from the triple point.  
Two supersonic shear layers are also evident: the outer one separates the wall-detached supersonic jet from the subsonic recirculating air entrained from the ambient; the inner shear layer, emanating from the triple point, separates the detached supersonic jet from the subsonic region downstream of the Mach disk.  Both shear layers impinge on the second shock waves system, where a second Mach disk can be seen just downstream of the nozzle exit. More detailed analysis of this TIC flowfield at $NPR=9$ is reported in Bakulu et al.\cite{Bakulu_jfm_21} and in Morisco et al.\cite{Morisco_23}.
Figures~\ref{fig:ddes_tic}b and c show the experimental and numerical mean wall-pressures and standard deviations: the position of the separation location is underpredicted by DDES, probably due to a grid induced separation effect, not completely contrasted by the shielding function~(\ref{eq:sh}). The standard deviation values are also underestimated, except towards the end of the nozzle; there is clearly a delay in the formation of the turbulent content when passing from the RANS model for the attached boundary layer to the LES model for the separated shear layer, the so called gray-area~\cite{spalart06}. Nevertheless, as we shall see in the following, the essential unsteady dynamics of the jet column is captured by the present simulation.

\begin{figure}[ht]
     \centering
          \subfigure[]{\includegraphics[width=0.45\textwidth]{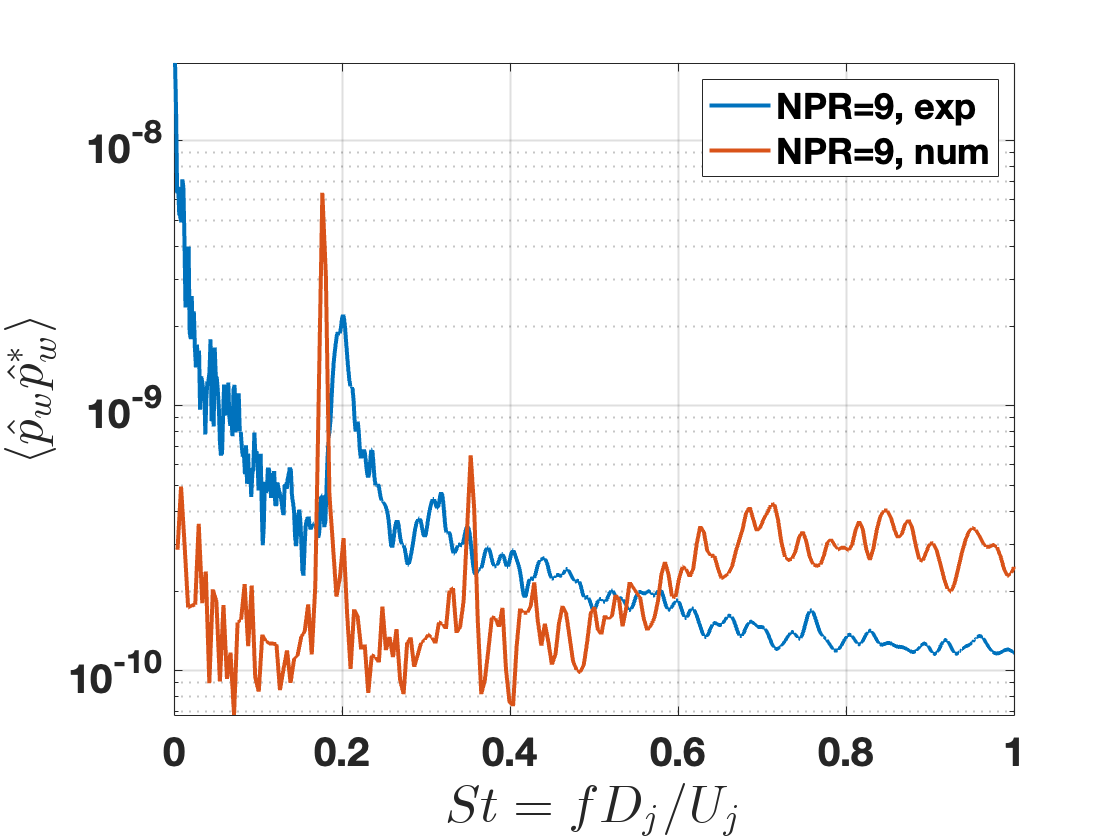}}
          \subfigure[]{\includegraphics[width=0.45\textwidth]{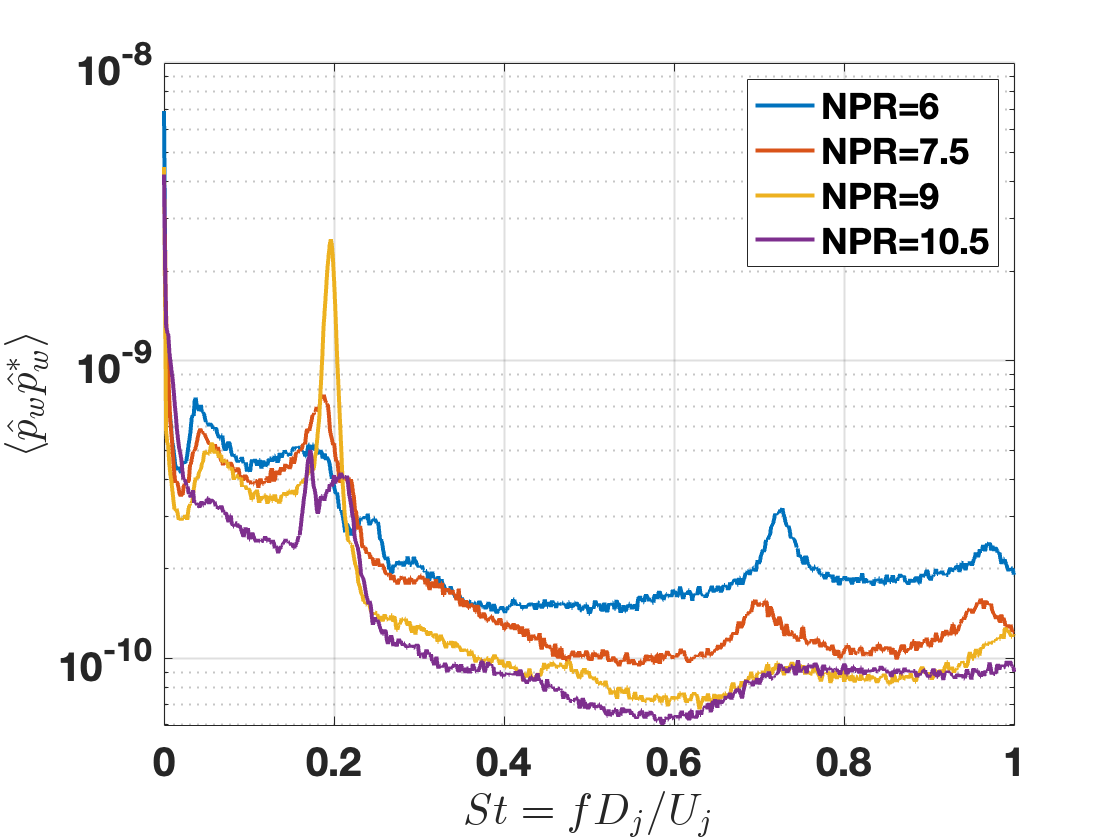}}
     \caption {a) PSD's of the experimental  and numerical  wall-pressure signals at $x/L=0.85$ for $NPR=9$ b) Experimental PSD's of the first pressure azimuthal mode as a function of frequency at $x/L = 0.67$ for $NPR=6, 7.5, 9, 10.5$.}
     \label{fig:PSD}
\end{figure}

We now present the spectral characterization of the experimental and numerical wall-pressure signals, and the static pressure is here always normalized with the stagnation chamber pressure $p_0$.
The power spectral density (PSD) of the pressure fluctuations is defined as the Fourier transform of their auto-correlation function:

\begin{equation}
    \left< \hat{p}_w \hat{p}*_w \right>(\omega) =\int_{-\infty}^{\infty} \left<p_w(t)p_w^*(t+\tau) \right> e^{-i\omega\tau}d\tau,
\end{equation}

 where $\hat{\bullet} (\omega)$ denotes the Fourier coefficient at the pulsation $\omega$ of a given time signal, $\left<\cdot\right>$ is the average operator, $\omega=2\pi f$ is the pulsation, and the asterisk indicates the complex conjugation. 
The PSD of the experimental wall-pressure signal extracted at $x/L=0.85$, $x/L=0$ being the nozzle throat, is reported in figure~\ref{fig:PSD}a together with the spectrum of the numerical signal extracted at the same location. 
As we already know from the literature~\cite{Bakulu_jfm_21}, we recognize two peak frequencies at $St=0.2$  and  $St=0.3$, where $St=f\cdot D_j/U_j$ and $D_j$ and $U_j$ are the fully expanded jet diameter and the fully expanded jet velocity~\cite{jaunet2017} , respectively. The first peak is correctly captured by the simulation, whereas the second peak is slightly over-predicted. We shall see that they are mostly contained in the first and second Fourier azimuthal modes.

Our focus is on the first azimuthal mode, since it is the only one responsible for the aerodynamic side loads. Therefore, the wall-pressure signals along all the nozzle wall are decomposed into Fourier azimuthal modes:

\begin{equation}
p_w,_m(x,t) = \int_0^{2\pi} p_w(x,\theta,t) e^{i m \theta}d\theta
\end{equation}

with the first three modes $m=0$, $m=1$ and $m=2$ indicating the breathing (symmetric), the helical (asymmetric) and the ovalization modes, respectively. Figure~\ref{fig:PSD}b  shows the power spectral densities of the asymmetric mode ($m=1$) of the experimental signals extracted  at $x/L = 0.67$ for  $NPR=6, 7.5, 9, 10.5$. It is evident that the PSDs of all the $NPRs$, except the first case, are characterized by remarkable peaks at $St\approx 0.2$,  with a predominance of the peak for the  $NPR=9$ test case. We know from literature that these peaks are coherent all along the nozzle wall where flow-separation occurs\cite{jaunet2017}.

\begin{figure}[ht]
     \centering
     \subfigure[]{
     \includegraphics[width=0.31\textwidth]{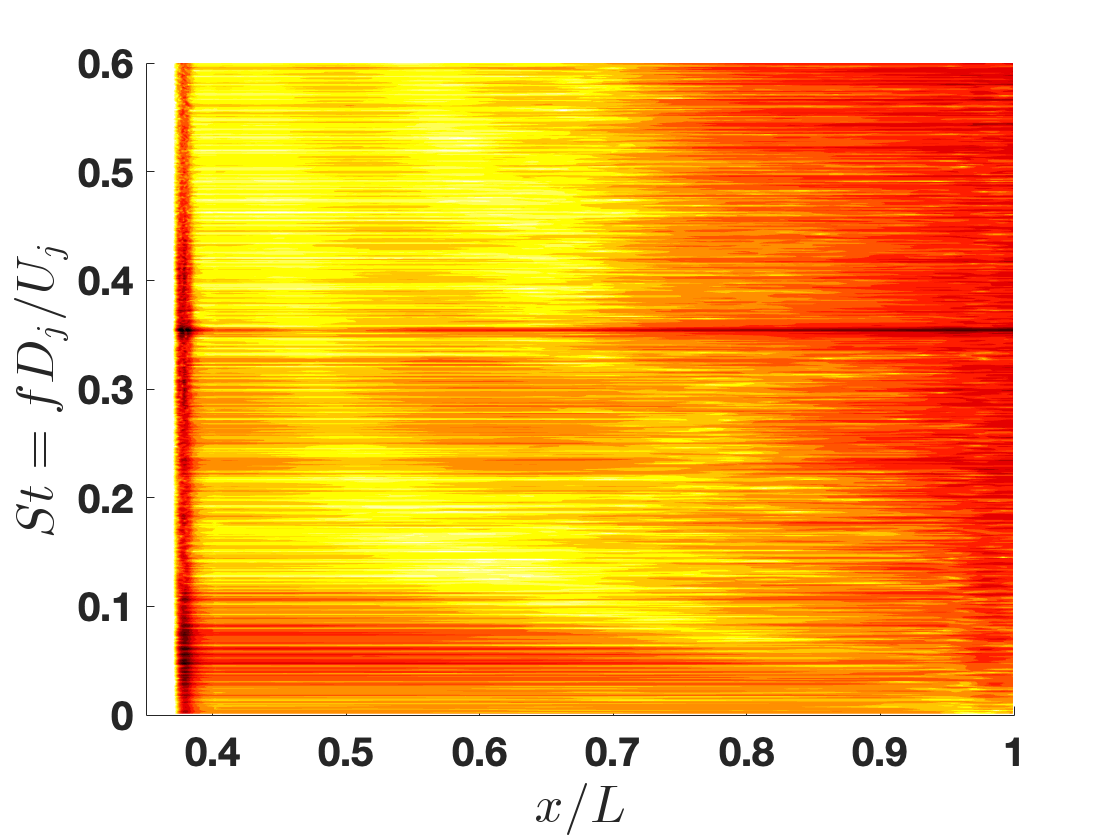}}
     \subfigure[]{
    \includegraphics[width=0.31\textwidth]{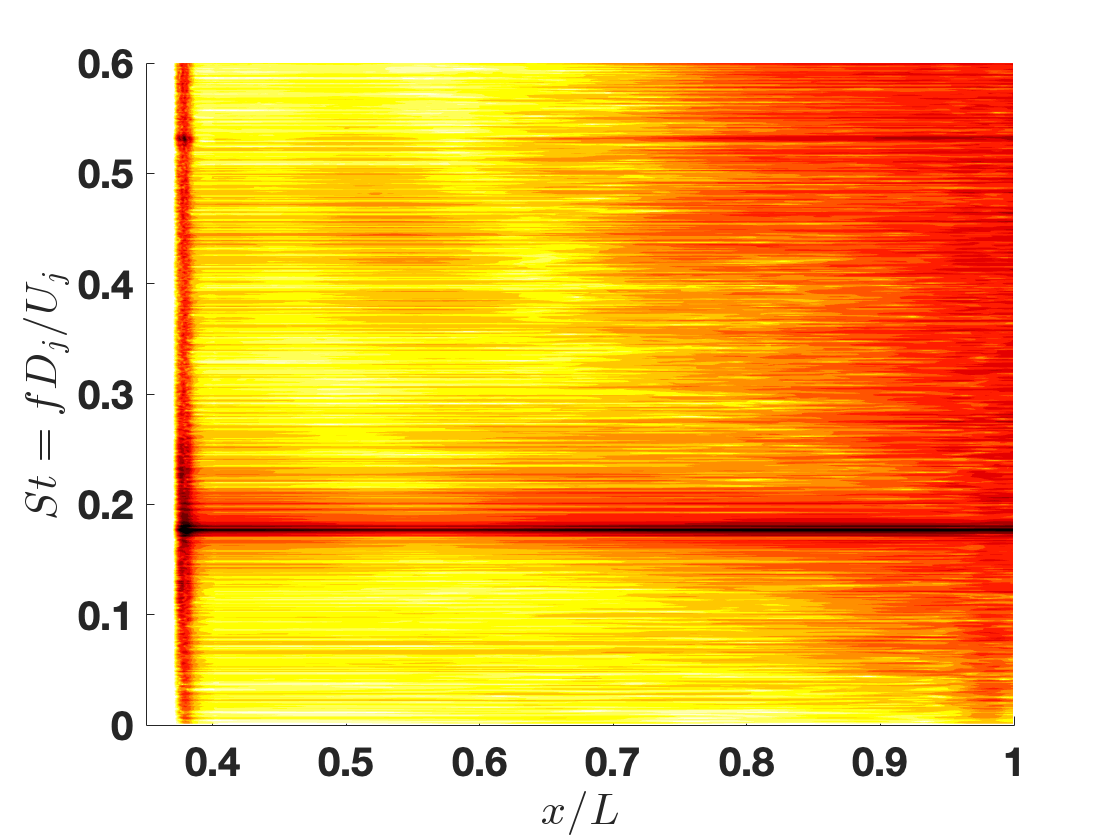}}
     \subfigure[]{
     \includegraphics[width=0.31\textwidth]{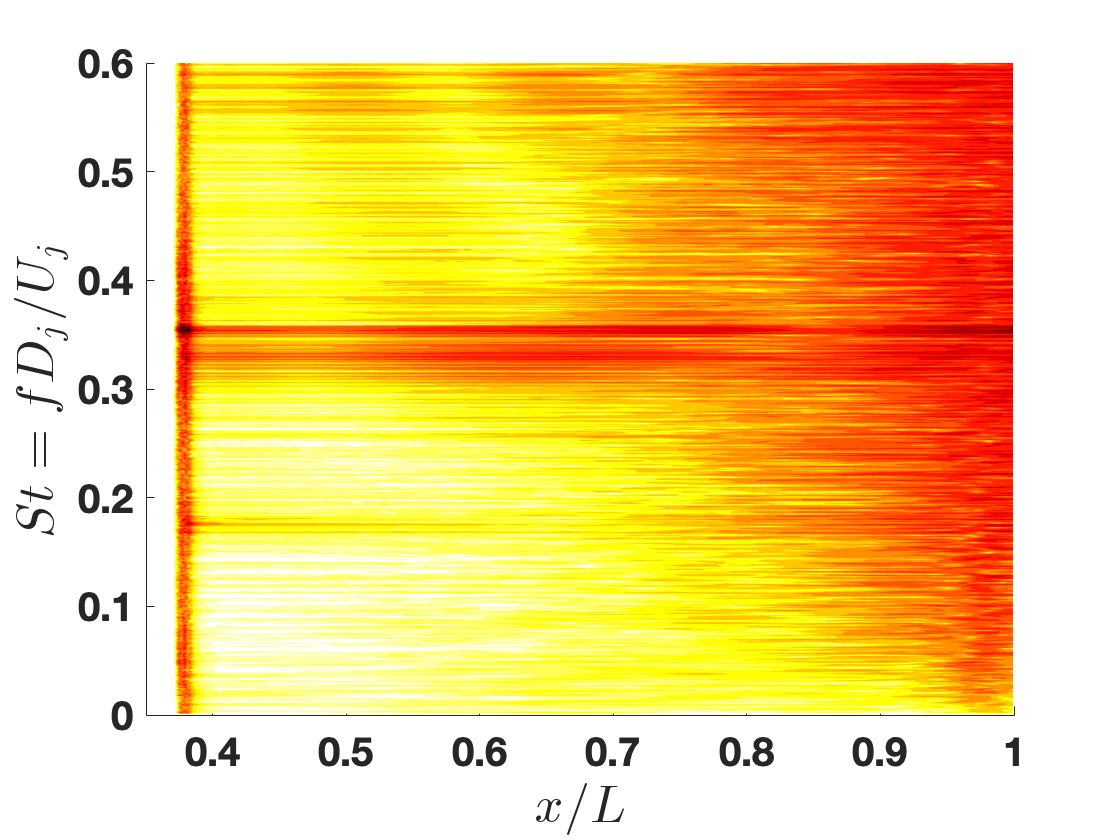}}
     \caption{Numerical PSD maps of the wall-pressure signals along the nozzle for the first three azimuthal modes at $NPR=9$: (a) $m = 0$, (b) $m = 1$, (c) $m = 2$.}
     \label{fig:PSDmap}
\end{figure}

The DDES simulation of the test case at $NPR=9$ confirms this important result. Figure~\ref{fig:PSDmap}, from a to c, shows the numerical power spectral density maps for the breathing, helical and ovalization modes all along the nozzle wall.
It is evident from this figure that the region corresponding to the separation point ($x/L\approx0.38$) is characterized by fluctuation energy spread over a wide range of frequencies, whereas downstream the separation region, each azimuthal mode is dominated by a particular temporal frequency, in good correspondence with the experimental peak frequencies. These signatures of the breathing, helical and ovalization modes persist all along the separated region confirming the observations of Jaunet et al.~\cite{jaunet2017} and Morisco et al.\cite{Morisco_23}. This seems to suggest that the entire jet column, at least the part inside the nozzle, is subject to unsteadiness at these frequencies.

\section{Wavelet analysis of intermittency}

Similarly to the approach adopted in Bernardini et al. \cite{bernardini2023unsteadiness}, the wavelet transform is applied here to extract the energetic intermittent events. Indeed, this approach allows for the extraction of local (in time) features that may be partially lost using Fourier analysis, thanks to the projection on a set of basis functions characterized by a compact support in both physical and frequency spaces~\cite{APracticalGuidetoWaveletAnalysis}. 
The wavelet transform is computed by the convolution of the wall-pressure signal $p_{w}(t)$ 
with the dilated (by the factor $k$) and translated (by the factor $t$) 
complex conjugate counterpart of a so-called mother wavelet, according to the following formalism:

\begin{equation}
w(k,t) = \frac{1}{\sqrt{k}} \int_{-\infty}^{+\infty} p_w(\tau)\Psi^{*}\left(\frac{\tau-t}{k}\right)\mathrm{d}\tau \,,
\label{eq:def}
\end{equation}

where $\Psi$ is the wavelet mother function, $k$ is a dilatation parameter indicating the time scale of the event under consideration, $t$ 
is the time-translation parameter.
A detailed theoretical framework can be found in Farge\cite{farge1992wavelet} and Mallat \cite{mallat1999wavelet}. In this study, the Morlet wavelet has been chosen, because it is a good compromise between time and frequency resolutions, compared to other mother functions. 

\begin{figure}[ht]
     \centering
          \subfigure[]{\includegraphics[width=0.45\textwidth]{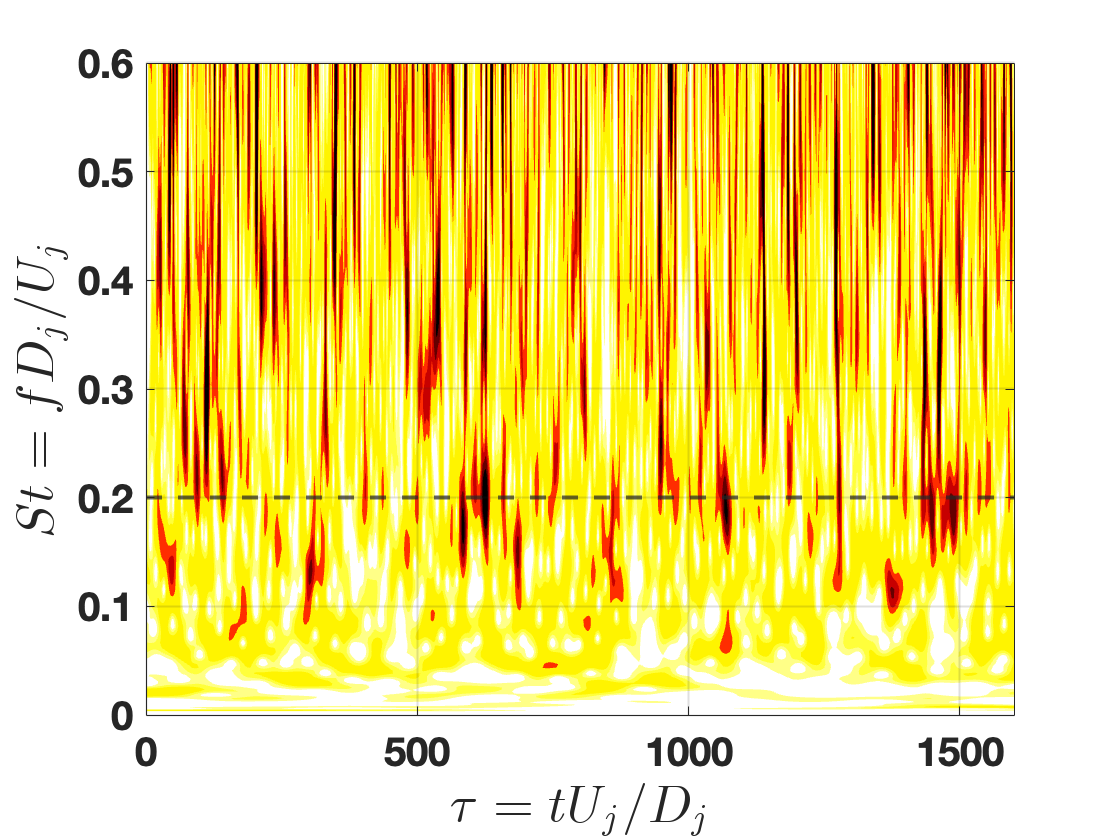}}
          \subfigure[]{\includegraphics[width=0.45\textwidth]{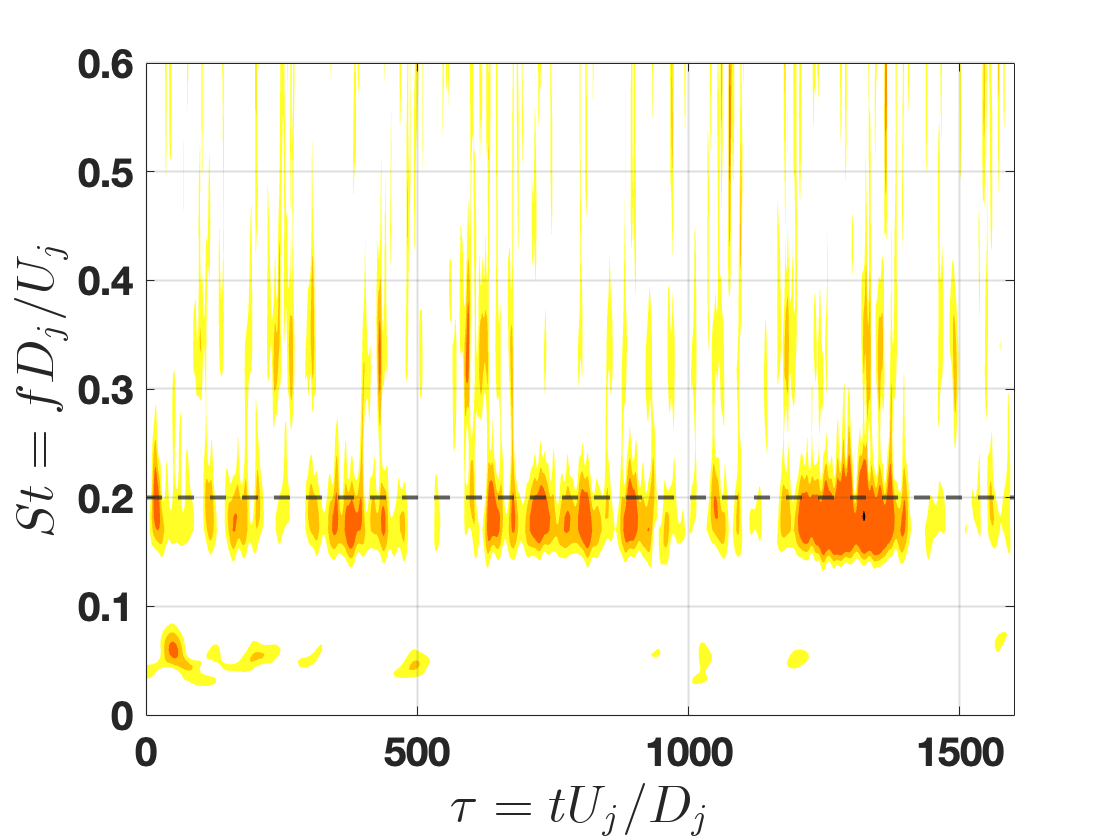}}
     \\
          \subfigure[]{\includegraphics[width=0.45\textwidth]{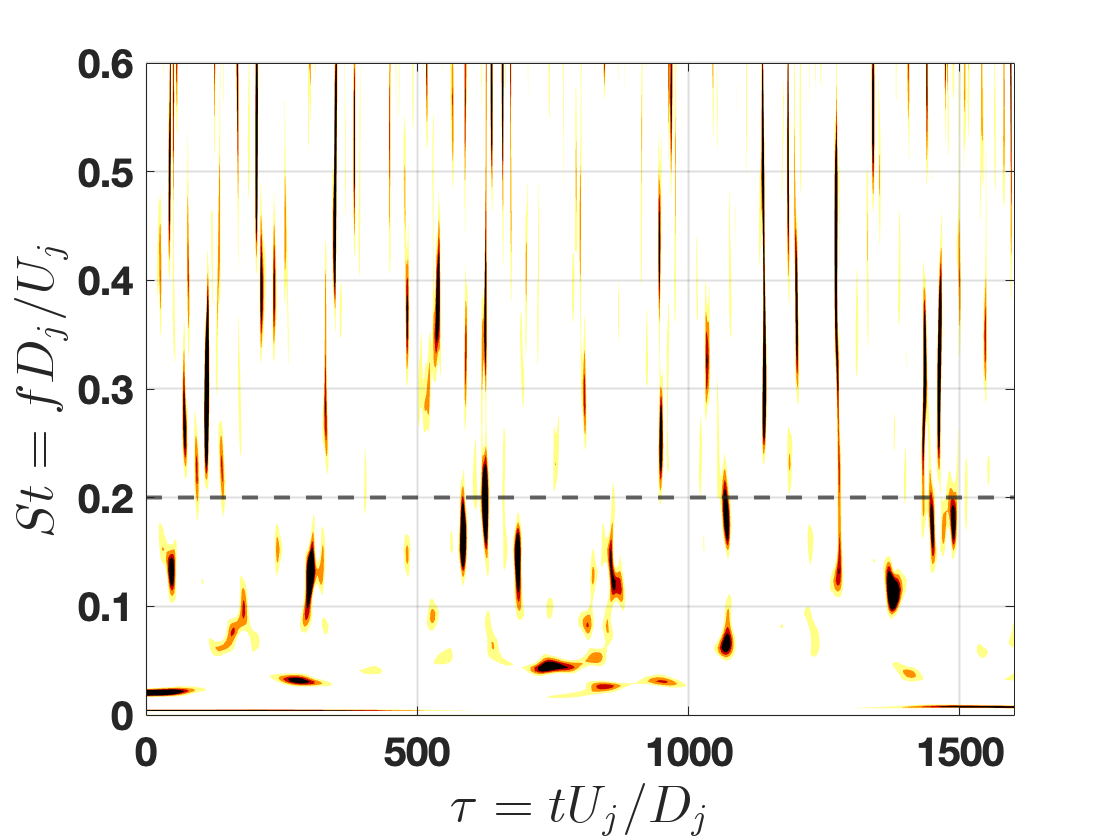}}
          \subfigure[]{\includegraphics[width=0.45\textwidth]{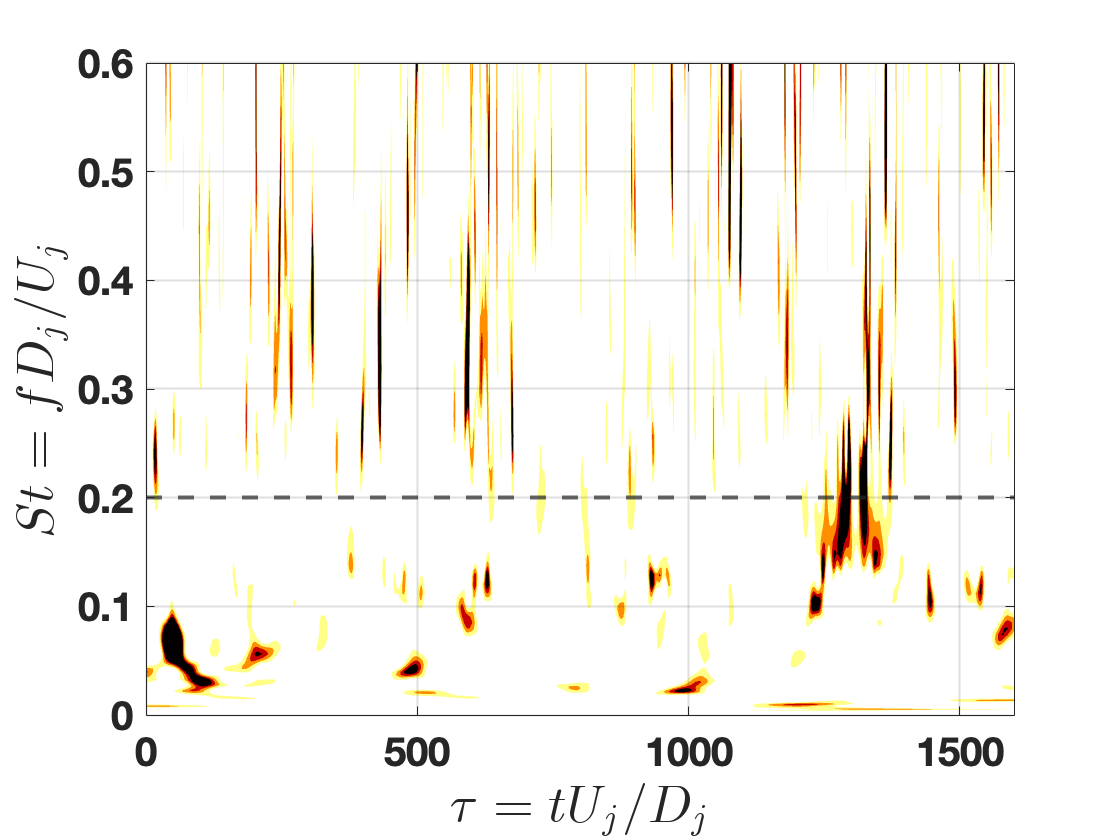}} 
     \label{fig:LIM2_num_exp}
     \caption{Scalograms of the experimental (a) and numerical (b) wall-pressure signals 
     and corresponding LIM2 maps (c, d) at $x/L=0.67$ for NPR$=9$. 
     The grey dashed lines indicate the characteristic frequency of the $m=1$ mode. 
     The LIM2 maps only show values greater than 3.}
     \label{fig:scalogram}
\end{figure}

The wavelet scalograms of the experimental  and numerical wall-pressure signals extracted at $x/L=0.67$ are reported in figure~\ref{fig:scalogram} (a and b respectively) for a segment of nondimensional time $\tau=t\cdot U_j/D_j = 1600$. It must be noted that the total time of analysis for the experimental test case is 60 seconds ($\tau \approx 5.89\cdot 10^5$), whereas the total integration time for the DDES is only of $\tau=1600$ due to the relevant computational resources required (despite the use of a hybrid RANS/LES method). Nevertheless, it is evident that the numerical simulation reproduces the qualitative time distribution of the events. 
We now focus on the frequency ranges corresponding to $m = 1$, that is around $St=0.2$ (dashed horizontal line in figure~\ref{fig:scalogram}): the wavelet decomposition highlights that the temporal evolution of the energy at this frequency  is strongly intermittent, since bursts of high energy are alternated to periods of relatively low energy levels. 

The qualitative aspects of the scalograms are very similar to those obtained by Camussi et al.~\cite{CAMUSSI20171} in their analysis of the near-pressure field of subsonic compressible jets. Therefore, in the following, we propose a 
procedure to select extreme events and a statistical analysis of these events inspired by their work.

Once the wavelet transform coefficients $w(k,\tau)$ are calculated , it is possible to obtain the scale-time distribution of the energy density $|w(k,\tau)|^2$ of the wall-pressure signals. Thanks to this property, an
effective indicator of intermittency is the squared local intermittency measure~\cite{camussi2021}, denoted as $LIM2$:

\begin{equation}
LIM2(k,\tau)=\frac{|w(k,\tau)|^4}{\langle|w(k,\tau)|^2\rangle_{t}^2}\,. 
\label{eq:LIM2}
\end{equation}

where $\langle \cdot \rangle_t$ indicates the time average of the quantity considered.
$LIM2$ can be interpreted as a time-scale-dependent measure of the flatness factor or kurtosis of a given signal. Therefore, the parameter $LIM2$ will be equal to 3 when the probability distribution is Gaussian, while the condition $LIM2>3$ identifies only those bursts of energy contributing to the deviation of the wavelet coefficients from a normal Gaussian distribution. 
The $LIM2$ maps are reported in figure~\ref{fig:scalogram} for the experimental (Fig.\ref{fig:scalogram}c) and numerical (Fig.\ref{fig:scalogram}d) signals extracted at $x/L=0.67$ for $NPR=9$, and we can immediately identify the bursts which induce a non-homogeneous distribution of energy in time.
Events characterised by large temporal scales (low-frequencies) are very rare, but also events with temporal scales (or frequencies) corresponding to the $m=1$ mode are rather sparse with respect to the turbulent bursts that are localized in a higher $St$ range. 

\begin{figure}[ht]
     \centering
     \includegraphics[width=0.45\textwidth]{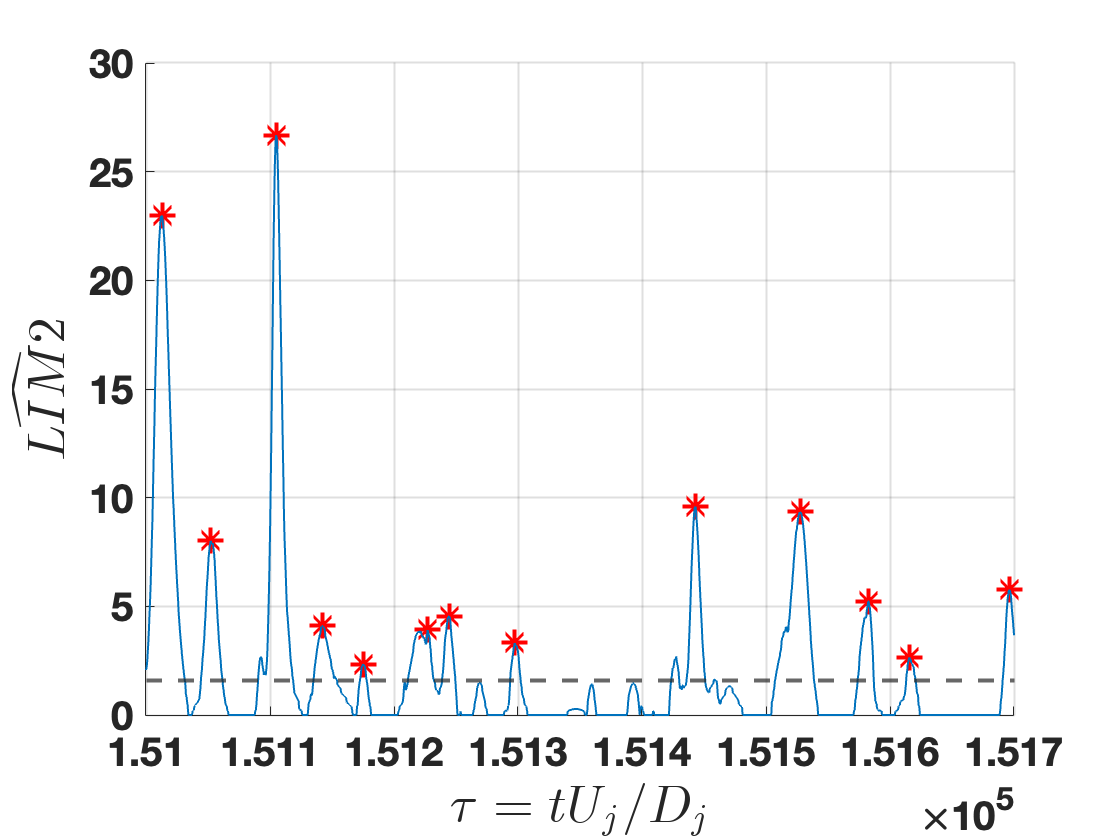} 
     \caption{Example of event selection in the experimental LIM2 map of the pressure signal at $x/L=0.67$ and $NPR$=9; $\widehat{LIM2}$ is the LIM2 averaged over the frequency range $0.0874 < St < 0.265$. the dashed line represents the threshold ($0.5\cdot\sigma_{LIM2}$) and the stars are the selected events.}
     \label{fig:LIM2-signal}
\end{figure}

Following the literature~\cite{camussi2021}, we investigate intermittency by evaluating  
the delay time $\Delta \tau=\Delta t \,U_j/D_j$ between the various energetic events. 
We focus our attention only on the events in the frequency range of the $m=1$ mode, since we want to characterize the aerodynamic loads given by the non-symmetrical azimuthal mode.
It should be noted that this intermittency is not related to the fluid dynamics concept of turbulence intermittency~\cite{kearney_2013}. Lacking a well-defined theoretical basis, an ad hoc definition of an \textit{``event''} is necessary, and to do so we mimic the methodology developed by Camussi et al.~\cite{camussi2021} based on the inspection of the LIM2 map.
Therefore, we compute the delay times according to the following tracking procedure:
\begin{enumerate}
    \item extraction of the LIM2 values corresponding to the scalogram region where $m=1$ dominates: 
    $0.0874 < St < 0.265$, 
    and average of the values in the selected frequency range; a short segment of the resulting time signal is shown in figure~\ref{fig:LIM2-signal}; 
    \item selection of times $t_i$ of the events with $LIM2>0.5 \cdot \sigma_{LIM2}$, 
    where $\sigma_{LIM2}$ is the standard deviation of the extracted LIM2 values;
    \item evaluation of $\Delta t_{m=1} (i) = t_{i+1} - t_i$;
    \item local average of the wavelet power spectrum $w^2$ in the selected frequency range; 
    evaluation of $w_i^2 = [w_i(t_i)]^2$ from the time distribution obtained.
\end{enumerate}

We report in figure~\ref{fig:LIM2-signal} a graphical example of the procedure. A sensitivity analysis has been preliminary conducted to test the robustness of the proposed procedure.

\section{Statistic analysis of events}
The purpose of this section is to investigate the nature of the extracted quantities through statistical analysis, to determine how these quantities vary with the $NPR$ and the distance along the nozzle wall from the separation location.

\subsection{Intermittency: time delay between events.}
\begin{figure}[ht]
     \centering
          \subfigure[]{\includegraphics[width=0.45\textwidth]{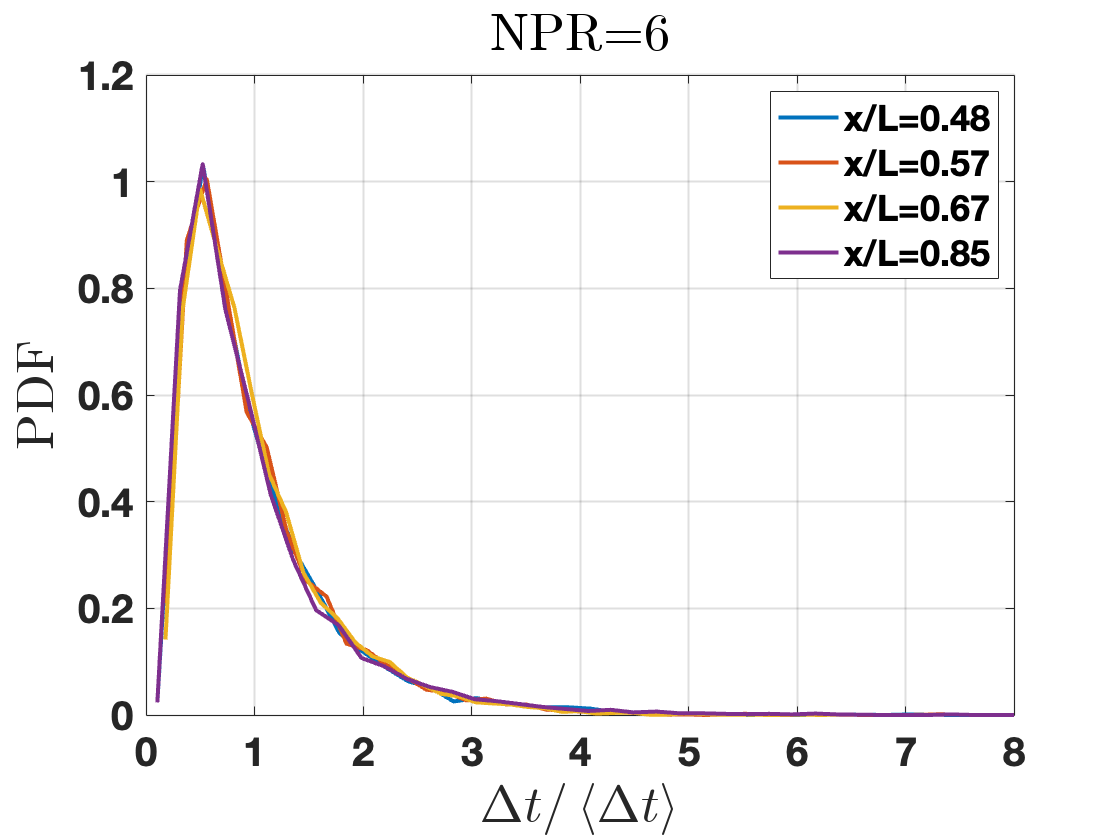}}
          \subfigure[]{\includegraphics[width=0.45\textwidth]{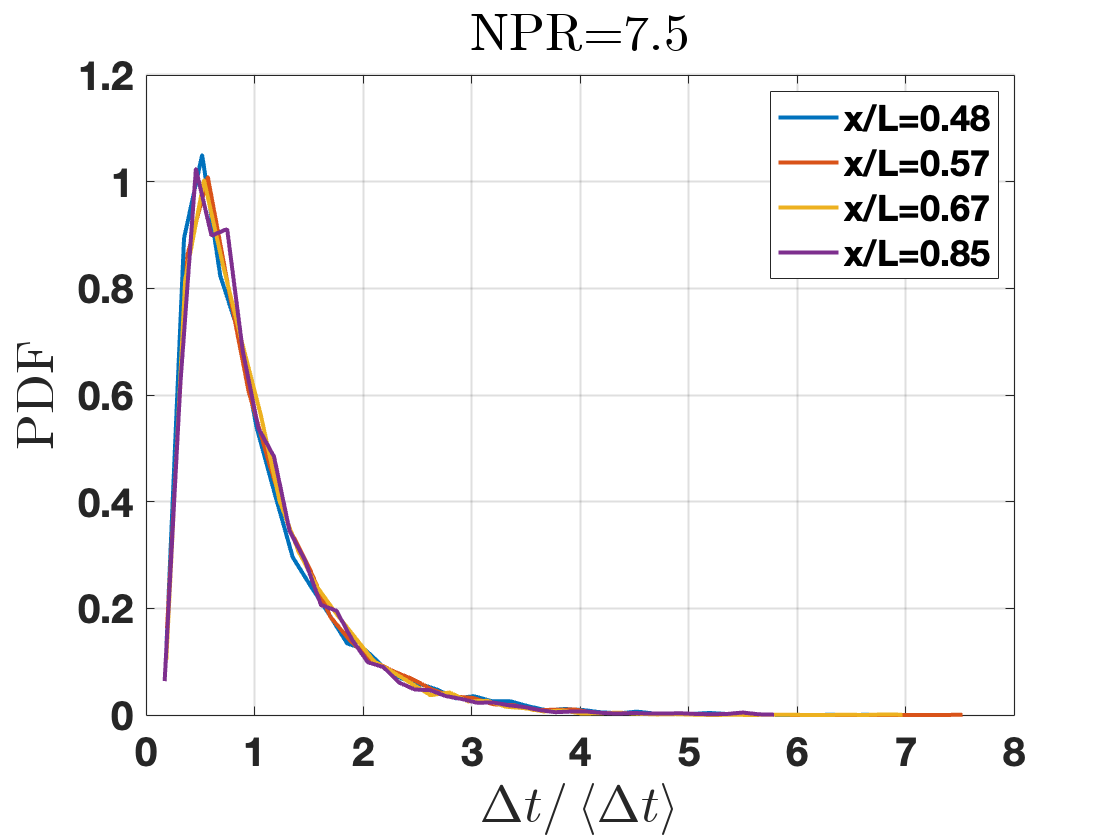}}
     \\
          \subfigure[]{\includegraphics[width=0.45\textwidth]{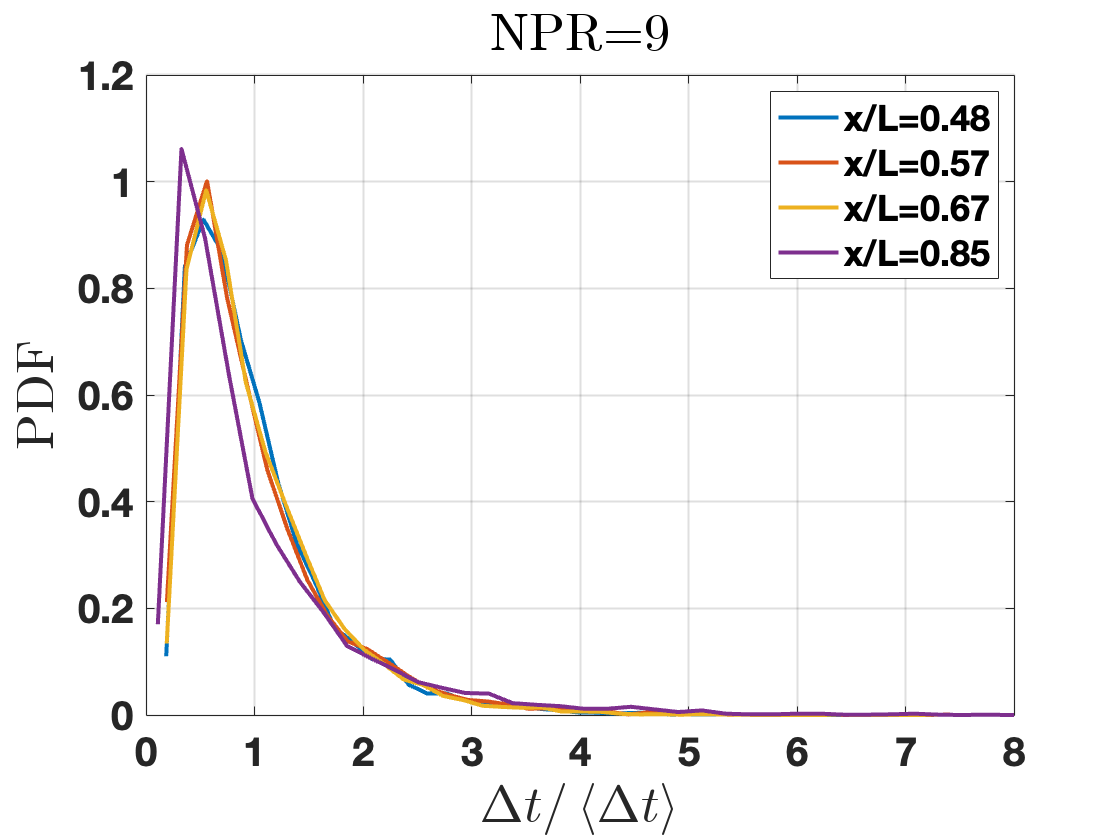}}
          \subfigure[]{\includegraphics[width=0.45\textwidth]{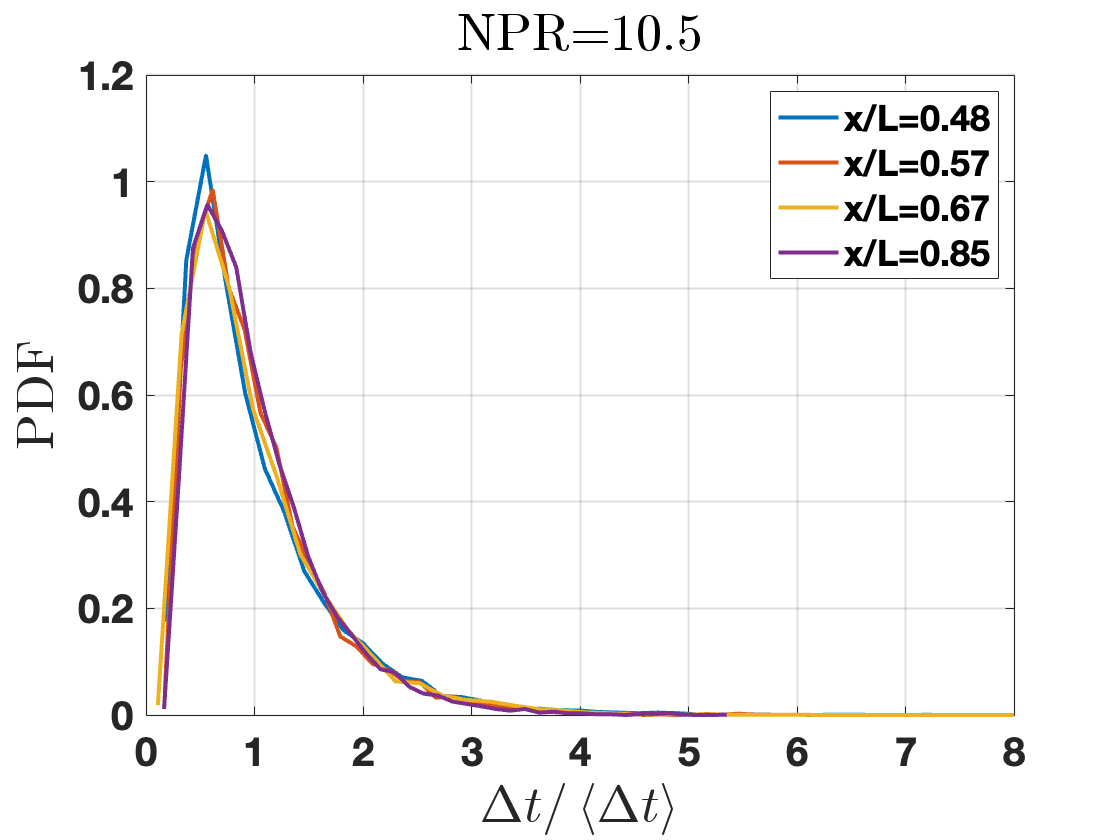}}
     \caption{PDF of the delay times between events for different distances from the separation location
             at top left) $NPR=6$; top right)  $NPR=7.5$; bottom left)  $NPR=9$; bottom right)  $NPR=10.5$.}
     \label{fig:pdf_m_all_npr}
\end{figure}

The \glspl{pdf} of the extracted time delays, normalized by their average value $\Delta t/\left< \Delta t \right>$ (or equivalently  $\Delta \tau/\left<\Delta \tau \right>$)
at $x/L=0.48, 0.67, 0.85$ and for all the $NPRs$ are shown in figure~\ref{fig:pdf_m_all_npr}. 
First, it appears that these statistical distributions share the same shape: they are almost independent of the location along the nozzle, indicating a global behaviour of the jet column, at least inside the nozzle. The second important aspect is that there is no obvious effect of the nozzle pressure ratio: the analysis of the \glspl{pdf} of the time delays along the nozzle wall for all the $NPR$s reveals the same qualitative aspects. 

\begin{figure}[ht]
     \centering
          \subfigure[]{\includegraphics[width=0.45\textwidth]{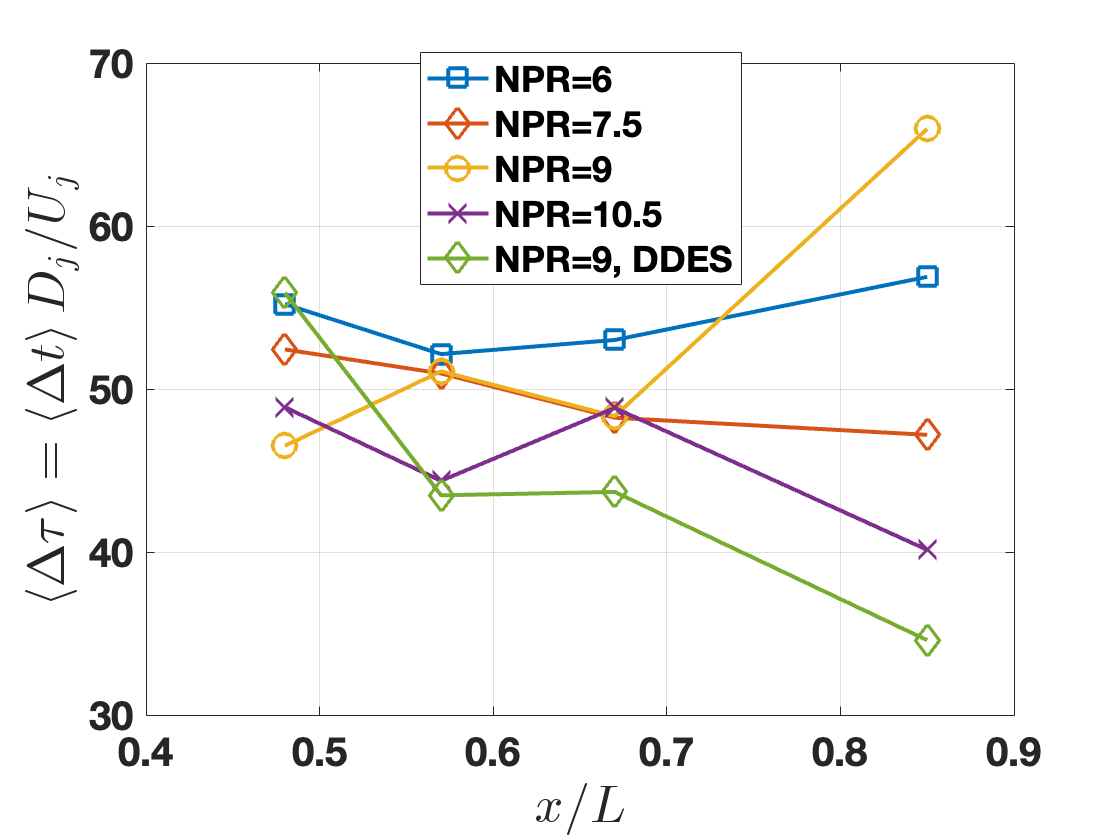}}
          \subfigure[]{\includegraphics[width=0.45\textwidth]{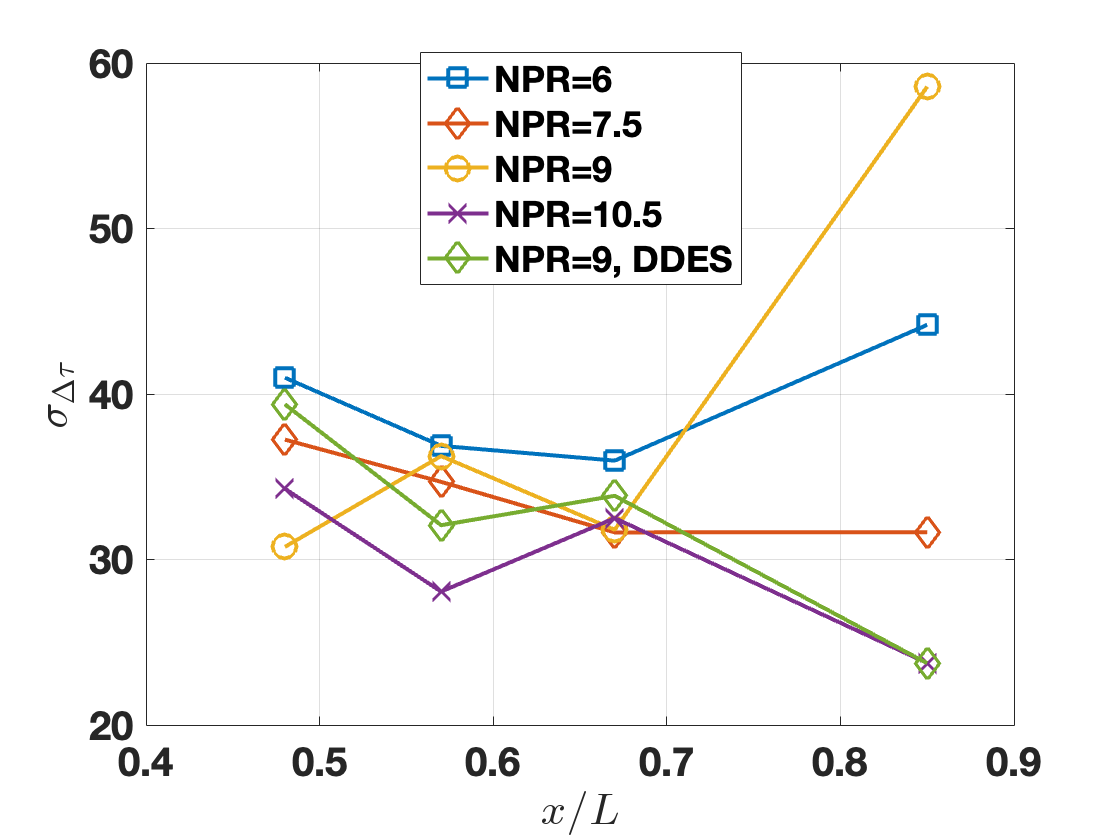}}
     \caption{Comparison of  $<\Delta \tau>$ (left) and of $\sigma_{\Delta \tau}$ (right)  for the different locations along the nozzle wall and for all the NPR's.}
     \label{fig:stats_m1_npr6_10}
\end{figure}

The behaviour of the nondimensional mean $\left<\Delta \tau \right>$ and the standard deviation $\sigma_{St}$ of the delay time along the nozzle wall are reported in figure~\ref{fig:stats_m1_npr6_10}. It appears that there is a small effect of the location $x/L$ and of the $NPR$ on both the mean values and the standard deviations, except at $x/L=0.85$ for $NPR=9$ where higher values are displayed. 
The figure reports also the mean delay times and the standard deviations evaluated from \gls{ddes}, since these are the only statistical quantities that we can capture, given the limited integration time. It can be seen that the values extracted from the numerical analysis are of the same order of magnitude of the experimental values. 

\begin{figure}[!h]
     \centering
          \subfigure[]{\includegraphics[width=0.45\textwidth]{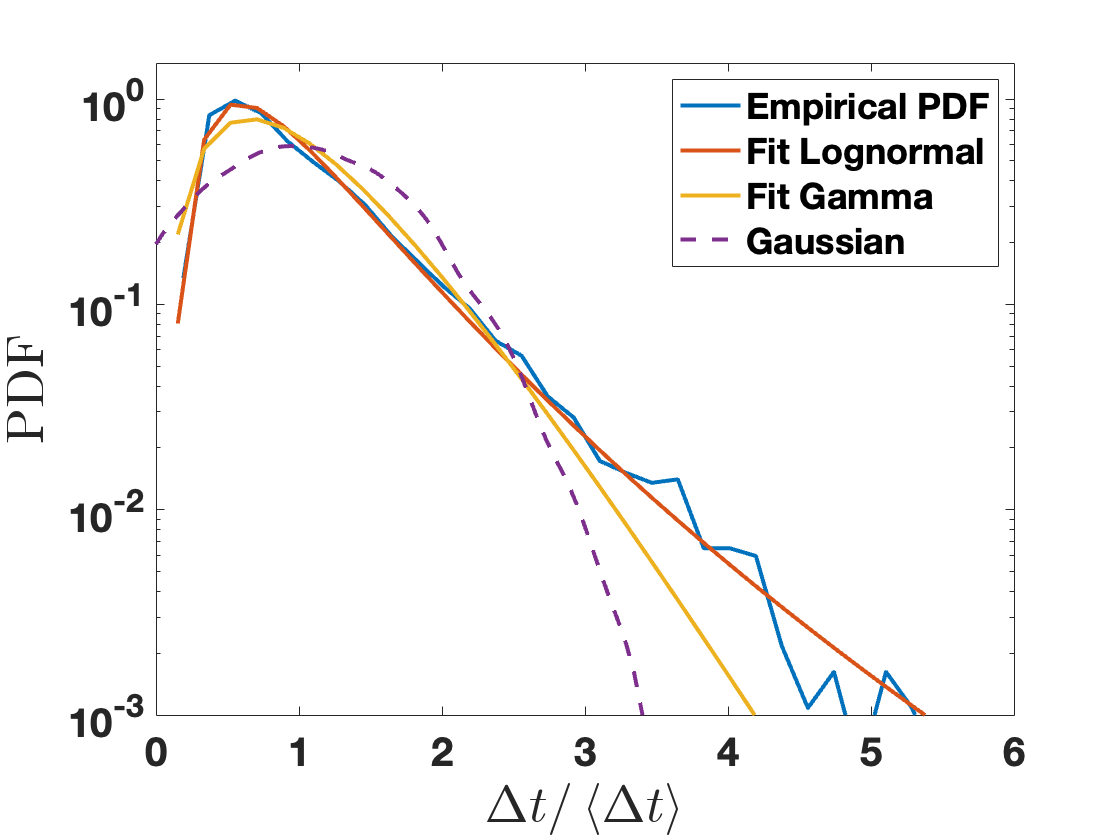}}
          \subfigure[]{\includegraphics[width=0.45\textwidth]{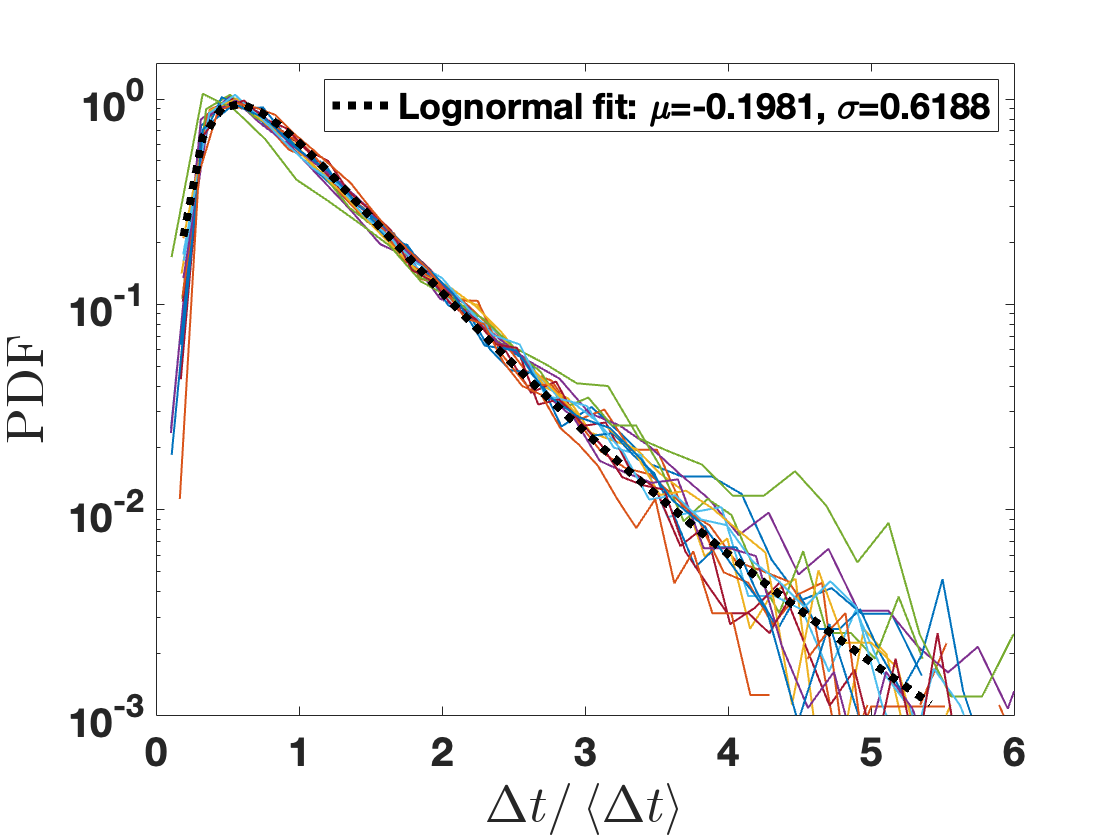}}
          \subfigure[]{\includegraphics[width=0.45\textwidth]{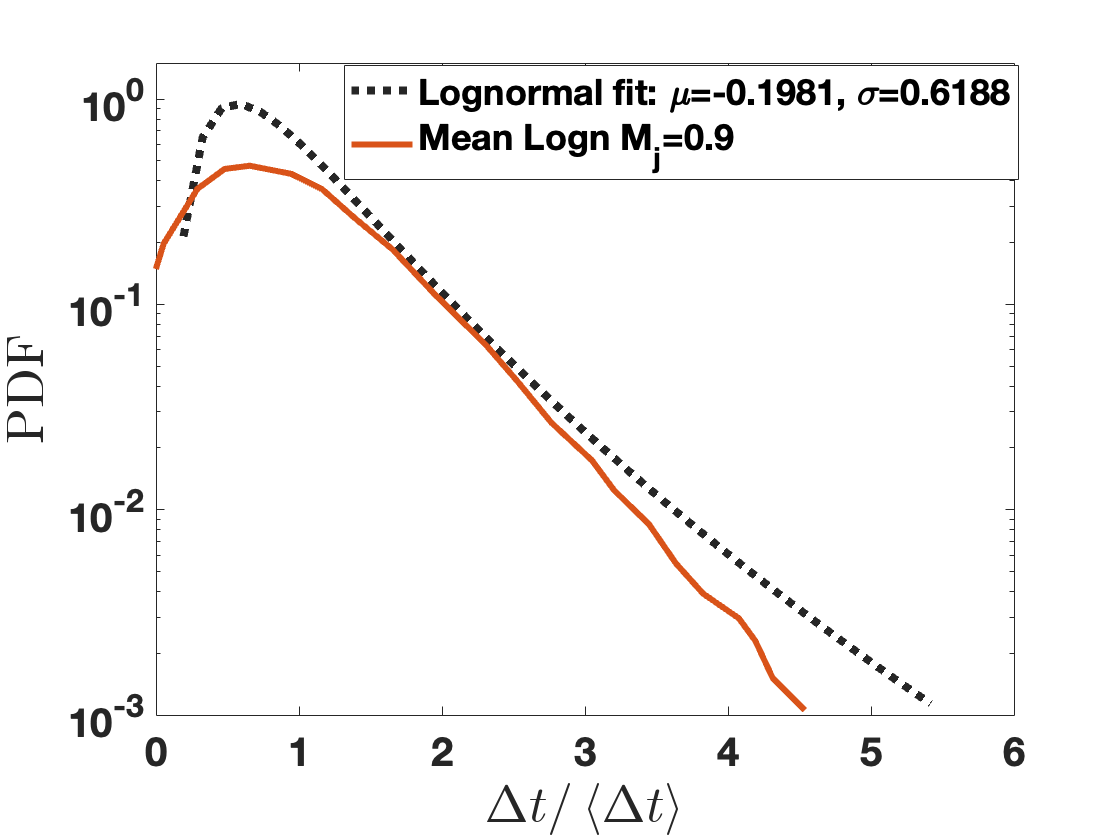}}
     \caption{a) Comparison between the empirical PDF ($x/L=0.67$, $NPR=9$), a log-normal fitting, a gamma fitting and a reference distribution for a Gaussian process; b) Intermittency distributions for all the NPRs and locations along the nozzle wall together with the proposed universal log-normal fit; c) comparison between the present unique log-normal fitting and the unique lognormal fitting of Camussi et al.~\cite{CAMUSSI20171}}. 
     \label{fig:lognorm_overall}
\end{figure}

Figure~\ref{fig:lognorm_overall}a shows the distribution of the time delays at $x/L=0.67$ for $NPR=9$ together with the $PDF$ of a reference Gaussian signal, obtained by a random number generator, a gamma fit and a log-normal fit. It is evident that the $PDF$ of the delay times is characterized by a heavy tail, a characteristics common to intermittent events~\cite{sapsis}. Then, we can see that the log-normal distribution is a better fit with respect to the gamma distribution (and other distributions according to a best fit test). The statistical distributions of all the test cases are then fitted with the log-normal distributions and the fitting parameters, location $\mu_{\Delta \tau/\langle \Delta \tau \rangle }$ and shape $\nu_{\Delta \tau/\langle \Delta \tau \rangle }$ are reported in table~\ref{tab:DT_parameters_table} together with the values of $\langle \Delta \tau \rangle $ and $\sigma_{\Delta \tau} $.
The \glspl{pdf} for all the 12 cases are shown in figure~\ref{fig:lognorm_overall}b, which shows a marked collapse of all the data onto a single curve. 
Therefore, a unique fitting distribution curve based on the averaged log-normal parameters ($\mu_{\Delta \tau/\langle \Delta \tau \rangle ,av}=-01981$ and $\nu_{\Delta \tau/\langle \Delta \tau \rangle ,av}=0.6188$) is added and shown as a black dashed line. It appears that the mean delay time is the controlling parameter of the distribution, and, once this mean value is known, the entire distribution can be well predicted by an universal log-normal distribution based on the average values of the location parameter $\mu_{\Delta \tau/\langle \Delta \tau \rangle ,av}$ and the shape parameter $\nu_{\Delta \tau/\langle \Delta \tau \rangle ,av}$.

The obtained results are in line with the findings of Camussi et al.\cite{CAMUSSI20171}: the statistical distributions of the delay times of noise-emitting events in the near field (hydrodynamic pressure) of two compressible subsonic jets at $M_{jet}=0.6$ and $M_{jet}=0.9$ were fitted with log-normal distributions. Figure~\ref{fig:lognorm_overall}c shows again the unique log-normal fitting of the present data together with the unique log-normal distribution of the delay times for the subsonic jet with  Mach number $M_j$ equal to 0.9 obtained by Camussi et al.~\cite{CAMUSSI20171}. It is interesting to note that the right tails of the log-normal fits are very close at the beginning, then the fitting for the supersonic flow data shows a heavier tail, indicating the higher probability of long intervals between different bursts. Following Donzis and Jagannathan~\cite{Donzis_2013}, we can expect density to follow a log-normal statistics. In fact, the continuity equation has a solution of the form $\rho(t)=\rho(0)\cdot e^{-\int_0^t \nabla \cdot \vec{u} d\tau}$ in Lagrangian coordinates, then if we assume long times the integral can be  considered as a sum of independent random variables. As a consequence, the central limit theorem would suggest a log-normal distribution for density. Since pressure and density fluctuations are related by the isentropic relation $p \sim \rho^{\gamma}$, then it should be easy to show that pressure fluctuations will also be log-normal.
Kearney et al.\cite{kearney_2013}, instead, investigated the noise-emitting events in the far field (acoustic pressure) of a data set of compressible subsonic jet, with jet Mach number ranging from 0.5 to 0.9, and they have found the best fitting for the delays times to be a gamma distribution.  This kind of distribution arises from processes that are of additive nature and the occurrences of the events are independent of each other.  

\begin{table}[ht]
    \centering
    \begin{tabular}{c c c c c c}
    \hline \hline
    NPR& x/L & $\langle \Delta \tau \rangle $ & $\sigma_{\Delta \tau} $ & $\mu_{\Delta \tau/\langle \Delta \tau \rangle }$ & $\nu_{\Delta \tau/\langle \Delta \tau \rangle }$ \\
        \hline
6 & 0.48 & 55.21   & 41.0 & -0.215 & 0.642 \\
6 & 0.57 & 52.1    & 36.9 & -0.202 & 0.626\\
6 & 0.67 & 53.0    & 36.0 & -0.194 & 0.618 \\
6 & 0.85 & 56.9    & 44.2 & -0.229 & 0.659\\
7.5 & 0.48 & 52.4  & 37.2 & -0.203 & 0.625 \\
7.5 & 0.57 & 51.0  & 34.7 & -0.190 & 0.607 \\
7.5 & 0.67 & 48.2  & 31.6 & -0.179 & 0.592 \\
7.5 & 0.85 & 47.2  & 31.6 & -0.181 & 0.592 \\
9 & 0.48 & 46.5    & 30.8 & -0.182 & 0.597 \\
9 & 0.57 & 51.1    & 36.2 & -0.204 & 0.630 \\
9 & 0.67 & 48.35   & 31.8 & -0.183 & 0.601 \\
9 & 0.85 & 66.0    & 58.6 & -0.299 & 0.757 \\
10.5 & 0.48 & 48.9 & 34.3 & -0.200 & 0.624 \\
10.5 & 0.57 & 44.4 & 28.1 & -0.169 & 0.577 \\
10.5 & 0.67 & 48.8 & 32.5 & -0.186 & 0.605 \\
10.5 & 0.85 & 40.1 & 23.7 & -0.152 & 0.549\\
\hline \hline
    \end{tabular}
    \caption{Parameters characterizing the log-normal PDF of the delay time $\Delta \tau/\langle \Delta \tau \rangle $.}
    \label{tab:DT_parameters_table}
\end{table}

\subsection{Event energy distribution}
\begin{figure}[!ht]
     \centering
          \subfigure[]{\includegraphics[width=0.45\textwidth]{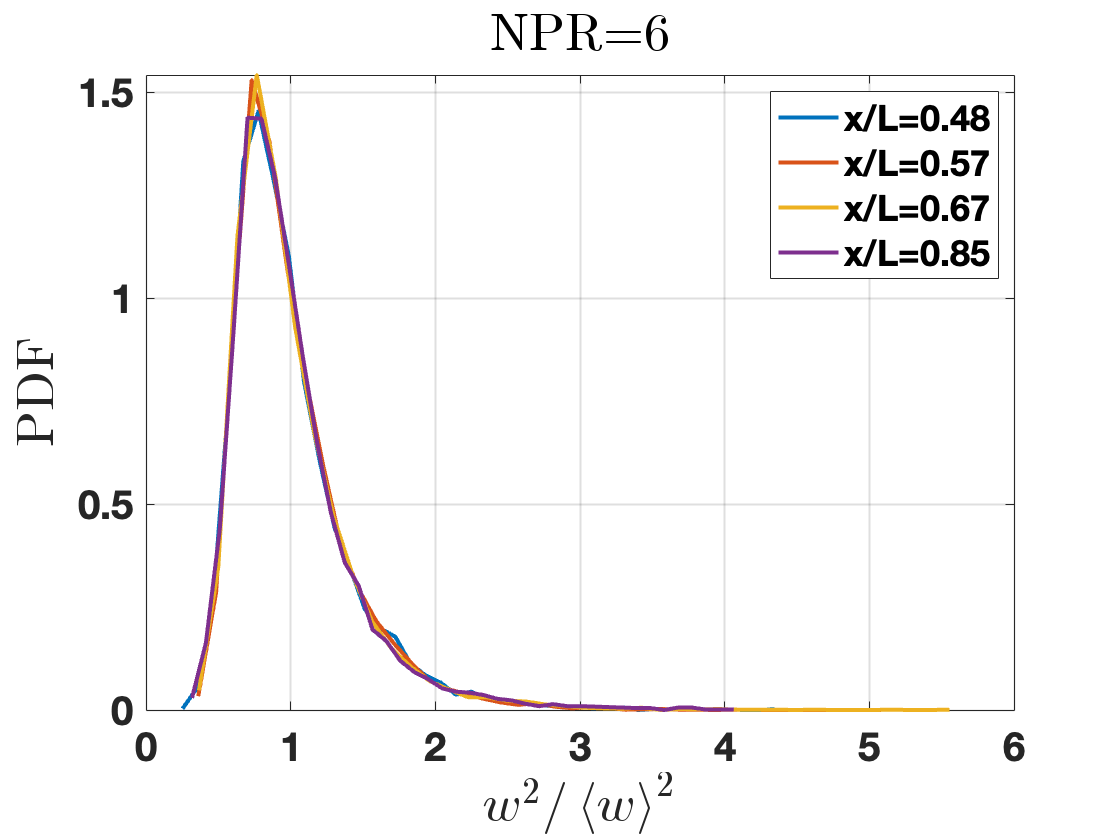}}
          \subfigure[]{\includegraphics[width=0.45\textwidth]{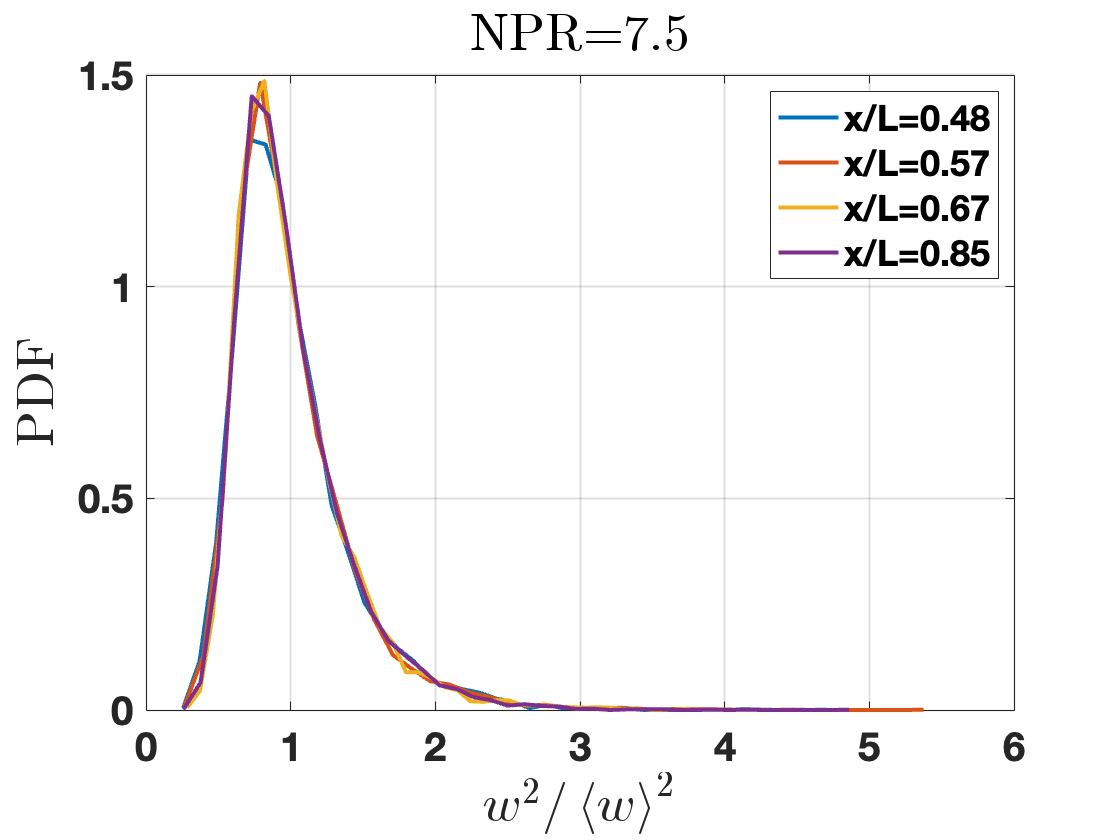}}
     \\
          \subfigure[]{\includegraphics[width=0.45\textwidth]{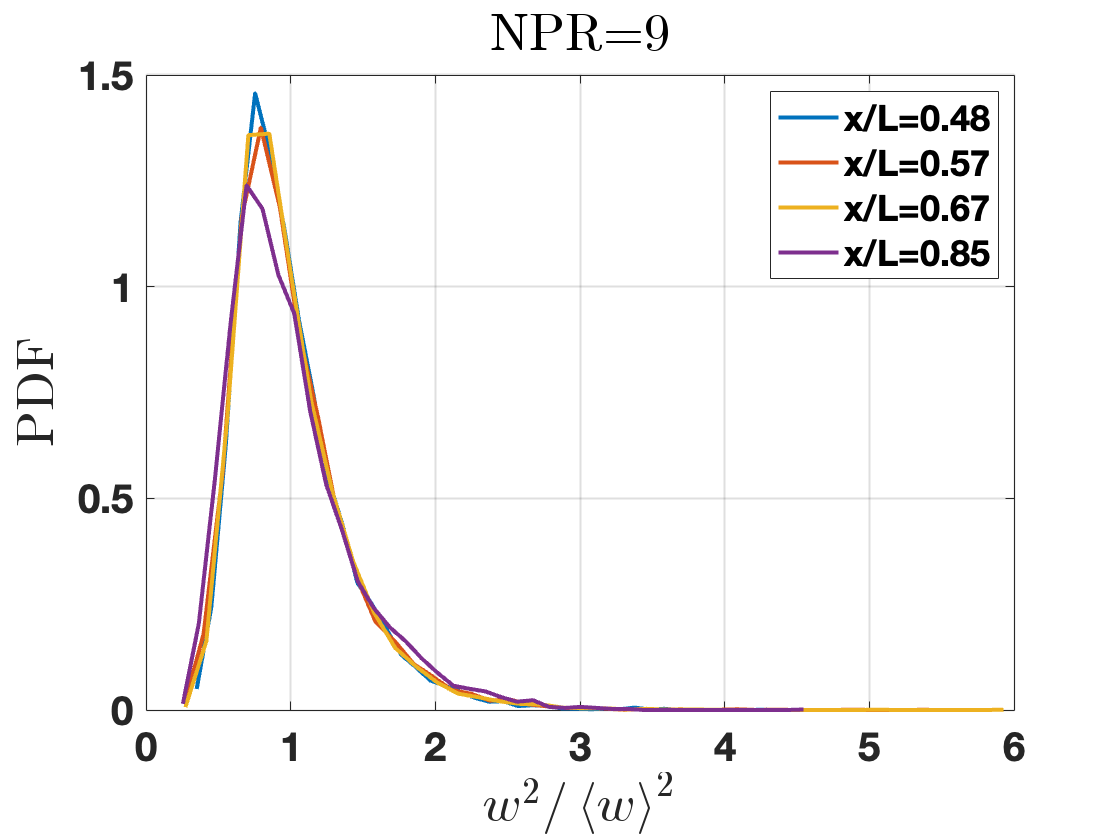}}
          \subfigure[]{\includegraphics[width=0.45\textwidth]{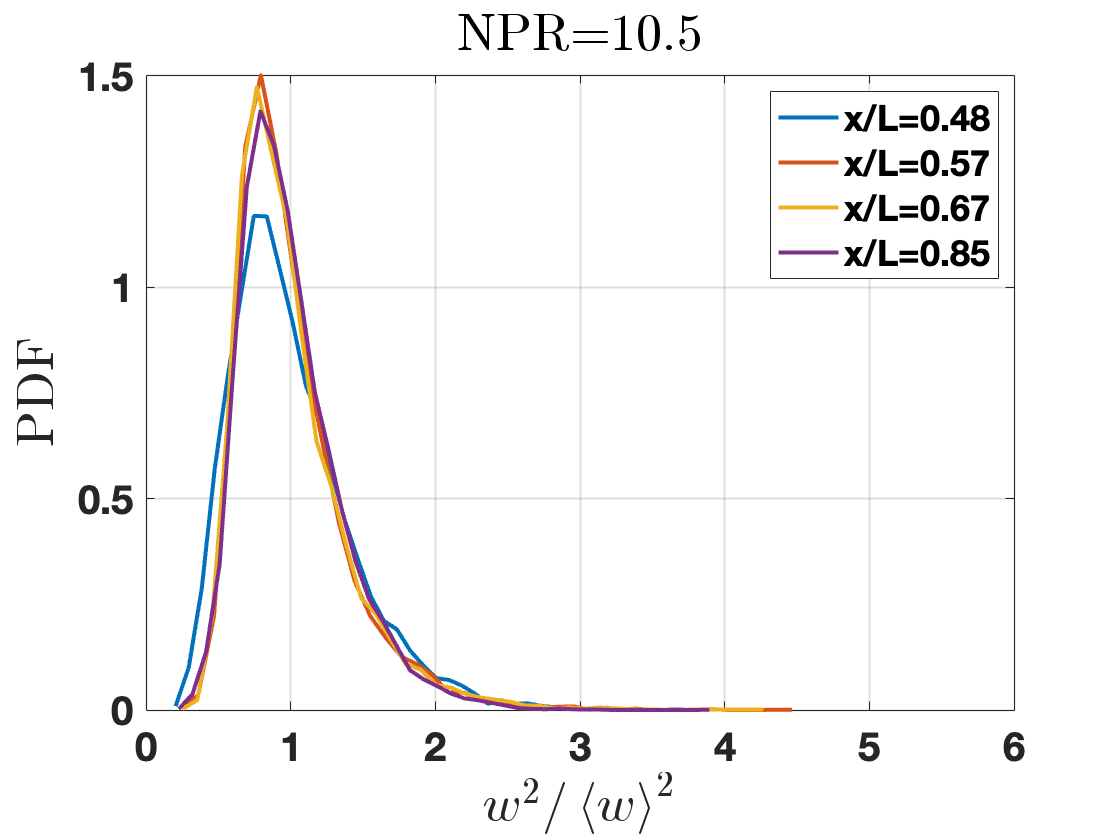}}
     \caption{PDF of the energy $w^2/\left<w^2\right>$ for the azimuthal mode $m_{1}$ for different distances from the separation location at top left) $NPR=6$; top right)  $NPR=7.5$; bottom left)  $NPR=9$; 
     bottom right)  $NPR=10.5$.}
     \label{fig:pdf_w2_all_npr}
\end{figure}

\begin{figure}[!ht]
     \centering
          \subfigure[]{\includegraphics[width=0.45\textwidth]{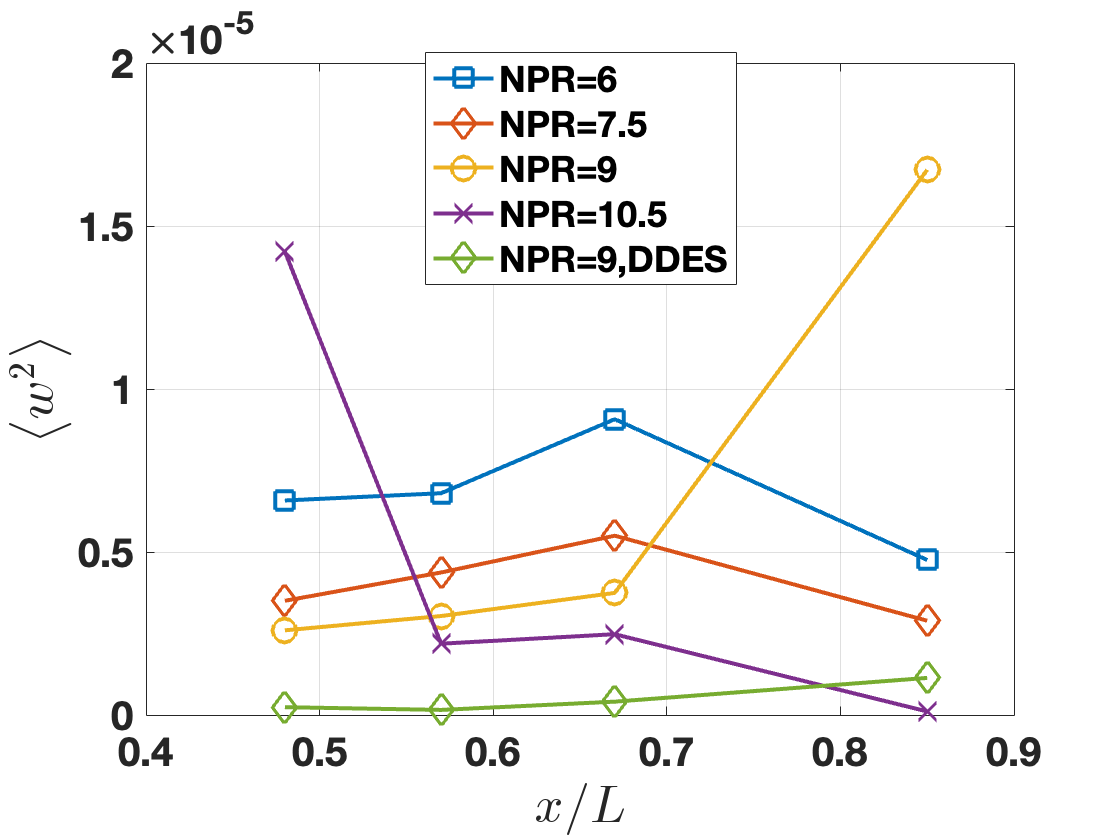}}
          \subfigure[]{\includegraphics[width=0.45\textwidth]{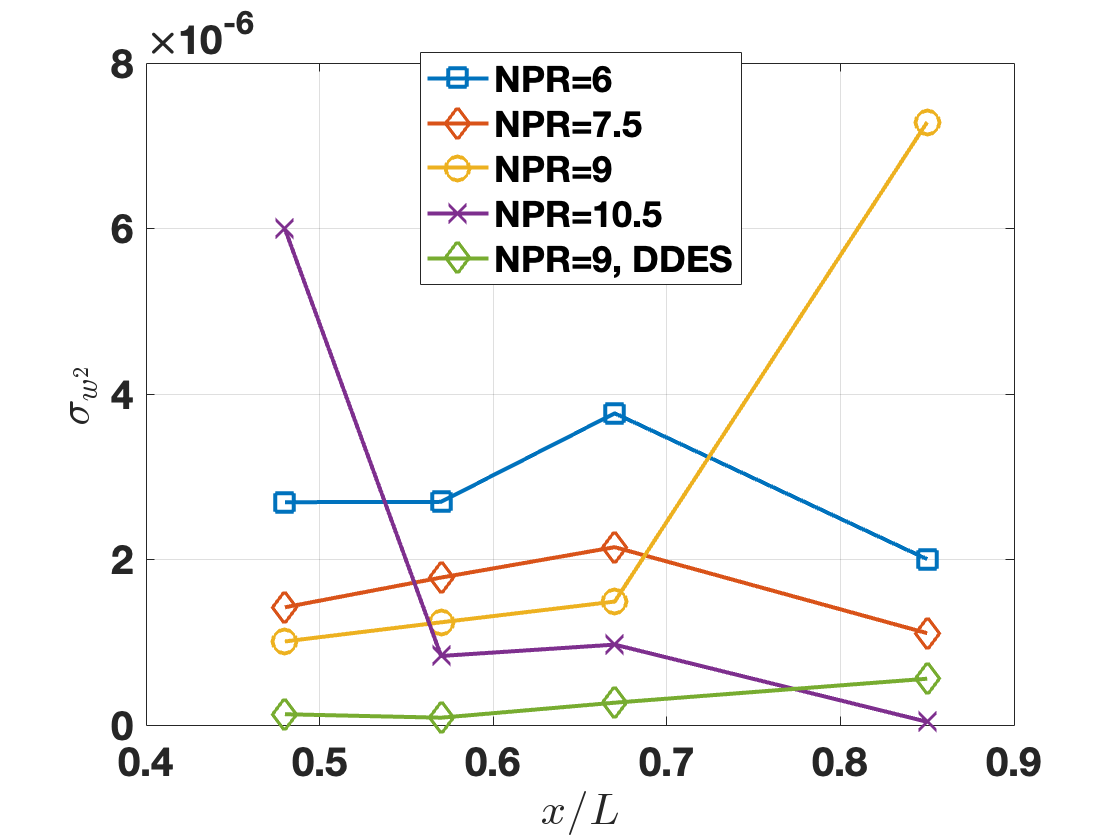}}
     \caption{Comparison of  $<w^2>$ (top left), of $\sigma_{w^2}$ (top right)  for the different locations along the nozzle wall and for all the NPRs.}
     \label{fig:stats_w2_npr6_10}
\end{figure}

\begin{figure}[!ht]
     \centering
          \subfigure[]{\includegraphics[width=0.45\textwidth]{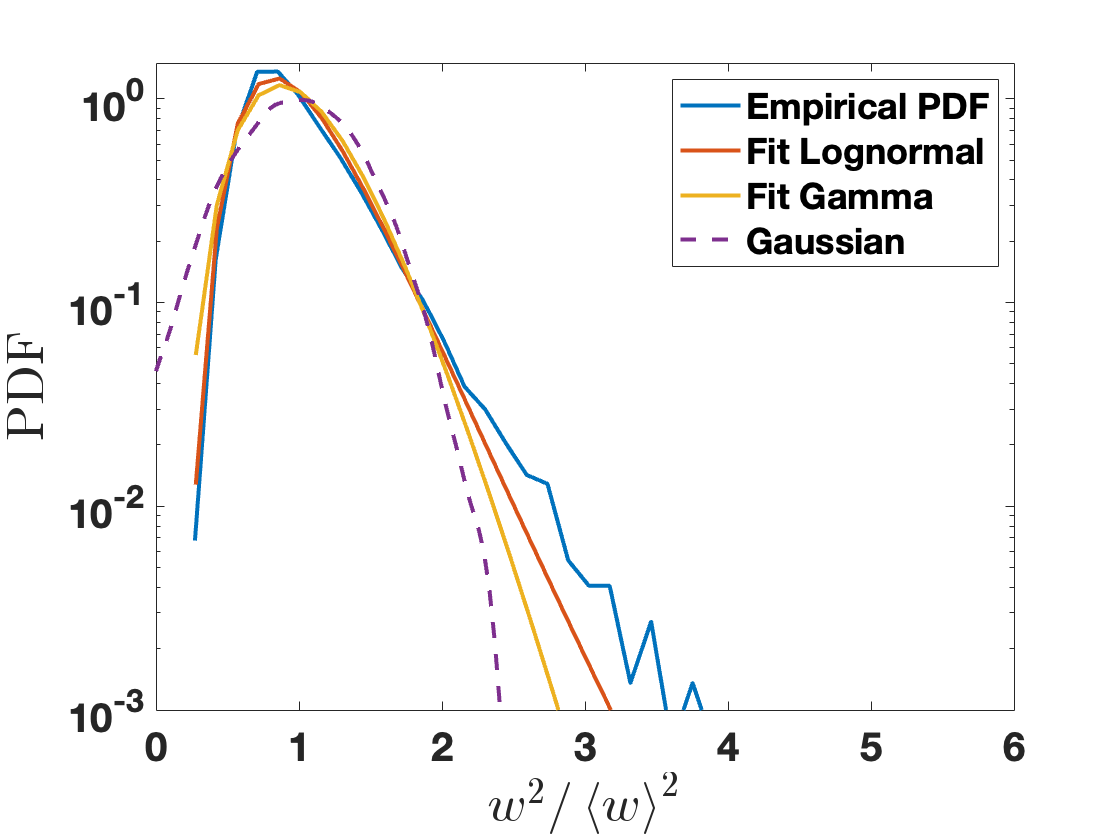}}
          \subfigure[]{\includegraphics[width=0.45\textwidth]{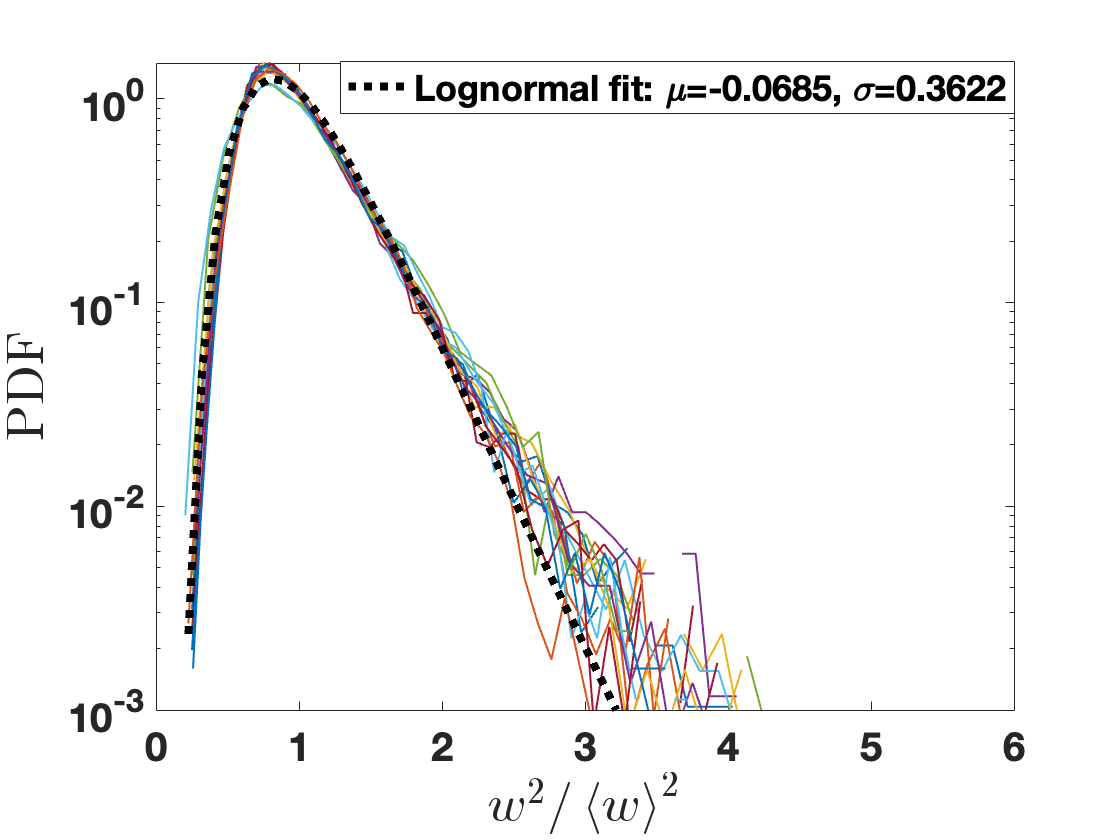}}
    \caption{Left) Comparison between the empirical PDF ($x/L=0.67$, $NPR=9$), a log-normal fitting, a $\gamma$ fitting and a reference Gaussian; right) amplitude distributions for all the NPRs and locations along the nozzle wall together with the proposed universal log-normal fit.} 
     \label{fig:lognorm_overall_wp}
\end{figure}

A similar approach has been adopted for the stochastic representation of the amplitude of the intermittent events. The PDFs of the selected-events' energy, defined as the square of the wavelet coefficients and normalized by their average values, $w^2/\langle w^2 \rangle$, are presented in figure \ref{fig:pdf_w2_all_npr} for all the analysed NPRs . These results suggest that both the position along the nozzle and the nozzle pressure ratio have a small influence on the distribution of the event energy. The only deviations can be appreciated in the last pressure sensor ($x/L=0.85$) for the case at $NPR=$9, as shown in figure~\ref{fig:pdf_w2_all_npr} (c), and in the first sensor ($x/L=0.48$) for the case at $NPR=$10.5, as shown in figure~\ref{fig:pdf_w2_all_npr} (d).
Figure~\ref{fig:stats_w2_npr6_10}  shows that both the average $\left<w^2\right>$ and the standard deviation $\sigma_{w^2}$ of the energy amplitude are almost independent of the distance and the $NPR$, with the relevant exceptions of the case  $NPR=9$ at  $x/L=0.85$ and of the case $NPR=10.5$ at $x/L=0.48$, as already noted in the analysis of the PDFs. In the latter case, $NPR=10.5$, the pressure sensor is very close to the separation shock, so it alternately records the pressure of the attached boundary layer and the separation zone. In the former case, $NPR=9$, we observe a constant increase in the values of $\left<w^2\right>$ moving along the nozzle, with a very high increment near the nozzle lip.
The values of $\left<w^2\right>$ and $\sigma_{w^2}$ obtained by the numerical simulation at $NPR=9$ are of the same order of magnitude as the experimental ones and also seem to follow their trend, even if with less intensity.
Figure~\ref{fig:lognorm_overall_wp}a shows the statistical distribution for the events selected at $x/L=0.67$ and $NPR=9$ and the best fit again resulted in the log-normal distribution. We also reported the gamma fitting and the reference Gaussian in order to highlight the tail behaviour. 
Figure~\ref{fig:lognorm_overall}b show all the log-normal fitting, whose parameters are reported in table~\ref{tab:W_parameters_table}, together with a unique fitting curve based on the averaged log-normal parameters ($\mu_{w^2/\left<w^2\right>,av}=-0.0685$ 
and $\nu_{w^2/\left<w^2\right>,av}=0.3622$). There is a very good collapsing of all the distributions with few exceptions. 

\begin{table}[ht]
    \centering
    \begin{tabular}{c c c c c c}
    \hline \hline
    NPR& x/L & $\langle w^2 \rangle $ & $\sigma_{w^2} $ & $\mu_{w^2}$ & $\nu_{w^2}$ \\
        \hline
6 & 0.48 & 6.58e-06 & 2.69e-06 & -0.0693 & 0.362 \\
6 & 0.57 & 6.83e-06 & 2.70e-06 & -0.0650 & 0.350 \\
6 & 0.67 & 9.07e-06 & 3.77e-06 & -0.0681 & 0.356 \\
6 & 0.85 & 4.76e-06 & 2.00e-06 & -0.0715 & 0.367 \\
7.5 & 0.48 & 3.50e-06 & 1.42e-06 & -0.0702 & 0.368 \\
7.5 & 0.57 & 4.37e-06 & 1.78e-06 & -0.0685 & 0.361 \\
7.5 & 0.67 & 5.50e-06 & 2.15e-06 & -0.0645 & 0.350 \\
7.5 & 0.85 & 2.89e-06 & 1.11e-06 & -0.0631 & 0.348 \\
9 & 0.48 & 2.60e-06 & 1.01e-06 & -0.0654 & 0.355 \\
9 & 0.57 & 3.03e-06 & 1.24e-06 & -0.0714 & 0.372 \\
9 & 0.67 & 3.74e-06 & 1.49e-06 & -0.0668 & 0.358 \\
9 & 0.85 & 1.67e-05 & 7.28e-06 & -0.0848 & 0.408 \\
10.5 & 0.48 & 1.42e-05 & 6.00e-06 & -0.0840 & 0.412 \\
10.5 & 0.57 & 2.19e-06 & 8.34e-07 & -0.0614 & 0.342 \\
10.5 & 0.67 & 2.48e-06 & 9.71e-07 & -0.0645 & 0.350 \\
10.5 & 0.85 & 1.09e-07 & 3.84e-08 & -0.0577 & 0.338 \\
\hline \hline
    \end{tabular}
    \caption{Parameters characterizing the log-normal PDF of the delay time $\Delta t$ (2.5M points)}
    \label{tab:W_parameters_table}
\end{table}

\section{Conclusions}

This work presents a combined experimental and numerical investigation into the unsteady dynamics of wall-pressure fluctuations in a supersonic over-expanded nozzle with free shock separation at different nozzle pressure ratios, with a specific focus on the statistical characterization of intermittent events associated with the main frequency of the first azimuthal mode. The test case at $NPR=9$ was reproduced by means of a delayed detached eddy simulation technique, which proved to be able to overall capture the global unsteadiness of the jet column and the main frequency of the first azimuthal mode ($St\approx 0.2$).
The energy distribution across frequencies of the wall-pressure oscillations and its evolution in time  have been tracked by means of the wavelet analysis, which is well suited to study unsteady phenomena.
It is observed that the first azimuthal mode is characterized by localized, sparse bursts of energy, alternated by periods of {\it silence}. The statistics of these events have been analyzed by selecting energetic occurrences from the wavelet scalogram within the frequency range associated with the first azimuthal mode for various $NPR$ values. Two key random variables were extracted: the time interval between successive bursts, \( \Delta \tau \), and the amplitude of their energy content, $w^2$, expressed in terms of the square of the wavelet coefficients.   
Statistical analysis reveals that the probability density functions of the variable $\Delta \tau $, when scaled by its average value, collapse into a almost universal log-normal distribution for all the NPRs tested and all the positions along the nozzle wall downstream of the separation line.  The main controlling parameter is therefore the mean intermittency $\left<\Delta t\right>$ (or $\left<\Delta \tau\right>$). Following Donzis and Jagannathan~\cite{Donzis_2013} and references therein, it may be argued that, for long times, the density fluctuations can be considered the result of the sum of independent random variables, and therefore, as a consequence of the central limit theorem, they must follow a log-normal distribution. Given the relation between density and pressure, the same hold true for the fluctuations of the latter.
As far as the amplitude of the events' energy (scaled by the corresponding average value) is concerned, a systematic comparison indicates that the log-normal distribution still offers  a consistent fit over the whole range of analysed test cases.  A universal behaviour can be again recovered when collecting all the distributions together.
These findings seem consistent with the results obtained in literature~\cite{kearney_2013, camussi2021} about the statistical characterization of the near and far pressure fields of subsonic compressible jets, whose noise is characterized by a high level of intermittency. In particular, it has been shown that all the distributions tend to collapse when appropriately scaled. In Camussi et al.~\cite{CAMUSSI20171}, where the near pressure field is investigated, the common distribution for the delay times is the log-normal one, whereas the amplitude distributions is better approximated by a gamma fitting. Kearney et al.~\cite{kearney_2013} analyzed the far pressure field and they found that the gamma fitting was the best approximation for both the delay times and the amplitudes. The physical reason behind this difference deserves more investigation. 
The statistical properties that characterize the delay times and the amplitudes of the events, together with the evidence that the aerodynamic load is only quasi-periodic at best, may form the basis of stochastic reduced-order models capable of capture the correct dynamics of the first azimuthal mode in nozzle over-expanded flow.

%
\section{Acknowledgments}
The simulation has been performed thanks to computational resources provided by the Cineca Italian Computing Center under the Italian Super Computing Resource Allocation initiative 
(ISCRA B/DESNON/HP10BLZA22). 

\bibliographystyle{elsarticle-num}
\bibliography{ExampleBiBTeXFile}

\end{document}